\pdfoutput=1
\documentclass[usenatbib]{mn2e}
\usepackage{apjfonts}
\usepackage{amssymb}
\usepackage{amsmath}
\usepackage{ctable}
\usepackage{fixltx2e} 

\newcommand{\be}{\begin{equation}}
\newcommand{\ee}{\end{equation}}

\newcommand{\etal}{et al.}

\newcommand{\msun}{M_{\sun}}

\newcommand{\paperone}{Paper {\small I}}
\newcommand{\papertwo}{Paper {\small II}}

\newcommand\plotone[2]
 {\centering \leavevmode \includegraphics[width={#2\columnwidth}]{#1}}
\newcommand{\plotside}[2]
 {\centering \leavevmode \includegraphics[width={#2\textwidth}]{#1}}
\newcommand{\acknowledgments}{\begin{small}\section*{Acknowledgments}\end{small}}
\newcommand\altaffilmark[1]{$^{#1}$}
\newcommand\altaffiltext[1]{$^{#1}$}
\title[Stellar Feedback, AGN Accretion, \&\ Obscuration]{Stellar and Quasar Feedback in Concert: Effects on AGN Accretion, Obscuration, and Outflows\vspace{-0.5cm}}

\vspace{-0.2cm}
\author[Hopkins \etal]{
\parbox[t]{\textwidth}{ 
Philip F.~Hopkins\thanks{E-mail:phopkins@caltech.edu}\altaffilmark{1},
Paul Torrey\altaffilmark{2}, 
Claude-Andr{\'e} Faucher-Gigu{\`e}re\altaffilmark{3}, 
Eliot Quataert\altaffilmark{4}, \&\ \\
Norman Murray\altaffilmark{5,6}
} 
\vspace*{6pt} \\
\altaffiltext{1}{TAPIR \&\ The Walter Burke Institute for Theoretical Physics, Mailcode 350-17, California Institute of Technology, Pasadena, CA 91125, USA} \\
\altaffiltext{2}{Department of Physics, MIT, 77 Massachusetts Avenue, Cambridge, MA 02139, USA} \\ 
\altaffiltext{3}{Department of Physics and Astronomy and CIERA, Northwestern University, 2145 Sheridan Road, Evanston, IL 60208, USA} \\ 
\altaffiltext{4}{Department of Astronomy and Theoretical Astrophysics Center, University of California Berkeley, Berkeley, CA 94720} \\
\altaffiltext{5}{Canadian Institute for Theoretical Astrophysics, 
60 St.\ George Street, University of Toronto, ON M5S 3H8, Canada} \\
\altaffiltext{6}{Canada Research Chair in Astrophysics} 
\vspace{-1.1cm}
}

\date{}
\begin{document}
\maketitle
\label{firstpage}

\vspace{-0.5cm}
\begin{abstract}
\vspace{-0.1cm}

We study the interaction of feedback from active galactic nuclei (AGN) and a multi-phase interstellar medium (ISM), in simulations including explicit stellar feedback, multi-phase cooling, accretion-disk winds, and Compton heating. We examine radii $\sim0.1-100\,$pc around a black hole (BH), where the accretion rate onto the BH is determined and where AGN-powered winds and radiation couple to the ISM. We conclude: (1) The BH accretion rate is determined by exchange of angular momentum between gas and stars in gravitational instabilities. This produces accretion rates $\sim0.03-1\,M_{\sun}\,{\rm yr^{-1}}$, sufficient to power luminous AGN. (2) The gas disk in the galactic nucleus undergoes an initial burst of star formation followed by several Myrs where stellar feedback suppresses the star formation rate (SFR). (3) AGN winds injected at small radii with momentum fluxes $\sim L_{\rm AGN}/c$ couple efficiently to the ISM and have dramatic effects on ISM properties within $\sim100\,$pc. AGN winds suppress the nuclear SFR by factors $\sim10-30$ and BH accretion rate by factors $\sim3-30$. They increase the outflow rate from the nucleus by factors $\sim10$, consistent with observational evidence for galaxy-scale AGN-driven outflows. (4) With AGN feedback, the predicted column density distribution to the BH is consistent with observations. Absent AGN feedback, the BH is isotropically obscured and there are not enough optically-thin sightlines to explain Type-I AGN. A `torus-like' geometry arises self-consistently as AGN feedback evacuates gas in polar regions.

\end{abstract}

\begin{keywords}
galaxies: formation --- galaxies: evolution --- galaxies: active --- 
star formation: general --- cosmology: theory
\vspace{-1.0cm}
\end{keywords}

\vspace{-1.1cm}
\section{Introduction}
\label{sec:intro}

The masses of super-massive black holes (BHs) correlate with various host galaxy bulge properties (\citealt{magorrian,FM00,Gebhardt00,hopkins:bhfp.obs,aller:mbh.esph,feoli:bhfp.1,kormendy:2011.bh.nodisk.corr}; for a review see \citealt{kormendy:2013.review.smbh.host.correlations}). The small scatter in these correlations \citep[relative to other galaxy properties;][]{hopkins:msigma.scatter,kormendy:2013.review.smbh.host.correlations}, together with constraints indicating that most BH mass is assembled in an optically bright quasar phase \citep{Soltan82,salucci:bhmf,yutremaine:bhmf,hopkins:old.age}, has led to the development of models where large-scale effects of feedback from accretion self-regulate BH growth at a critical mass \citep{silkrees:msigma,king:msigma.superfb.1,dimatteo:msigma,murray:momentum.winds}. Gas inflows triggered by some process fuel rapid BH growth in a nuclear starburst \citep{diamondstanic:2012.agn.fueling.in.starburst.cusps,mushotzky:2014.agn.present.in.high.sfr.cusps}, until feedback begins to expel nearby gas and dust. This ``blowout'' results in a short-lived, bright optical quasar that, having expelled its fuel supply, fades and leaves a remnant on the observed BH-host correlations \citep{hopkins:lifetimes.methods,hopkins:lifetimes.obscuration}. This general scenario has been able to explain many quasar observables, including luminosity functions, lifetimes, and BH mass functions \citep{hopkins:lifetimes.interp,hopkins:merger.lfs,hopkins:groups.qso,hopkins:seyfert.bimodality,volonteri:xray.counts,menci:sam,somerville:new.sam,lapi:qlf.sam,tortora:2009.agn.jet.fb.and.ell.colors}. It has also been speculated that this feedback might ultimately have a large impact throughout the AGN host galaxy, expelling or heating gas and explaining the rapid quenching of star formation in massive galaxies \citep{granato:sam,scannapieco:sam,croton:sam,hopkins:groups.ell,antonuccio-delogu:2008.jet.fb.destroying.sf.clouds}, and considerable observational evidence has emerged for this in recent years \citep[see e.g.][]{nesvadba:2010.maintenance.feedback.suppressing.sf.efficiency,cimatti:2013.agn.outflow.hundreds.kms.common.intermediate.ssfr.highz,lamassa:2013.low.ssfr.in.agn.hosts.lowz,shimizu:2015.agn.specific.sfr.anti.correlated.lowz,guillard:2015.agn.driven.turbulence.as.sf.suppressant,alatalo:2015.ngc.1266.outflow.suppressing.sf.via.turbulent.injection}.

High-velocity outflows can be driven from the BH accretion disk by a variety of physical processes including, e.g.,  radiation pressure on lines and dust, magnetic processes, or Compton heating \citep[see e.g.][]{blandfordpayne:mhd.jets,begelman:agn.compton.heating,chang:qso.rad.feedback,sanders88:quasars,koniglkartje:disk.winds, murray:1995.acc.disk.rad.winds,elvis:outflow.model,proga:disk.winds.2000,proga:disk.winds,silk:msigma.superfb.sb,murray:momentum.winds,batcheldor:outflow.mechanism,tortora:2009.agn.jet.fb.and.ell.colors}.   These manifest themselves observationally as ultra-fast outflows \citep[e.g.]{tombesi:2010.ufo.outflows,tombesi:2013.ufo.warm.absorbers.may.be.connected.through.outflow,tombesi:2015.ufo.molecular.outflow.same.galaxy}, the broad emission line regions and broad absorption line quasars \citep[e.g.][]{weymann:BALs,dekool:large.outflow.1,gabel:large.outflow,ganguly:qso.outflows}, more moderate velocity outflows ($v\sim10^{2}-10^{3}\,{\rm km\,s^{-1}}$) associated with the narrow line region  \citep{laor:warm.absorber,crenshaw:nlr,Steenbrugge:outflow.mdot,krongold:warm.absorber.outflow.rate}, as well as  quasar absorption and occultation systems \citep[e.g.][]{mckernan:1998.agn.occultation.by.clumpy.outflow,turner:2008.clumpy.agn.disk.wind,miller:2008.clumpy.agn.disk.wind}.   Observations on {\em galaxy} scales have also provided strong evidence for powerful molecular, atomic, and ionized outflows with velocities $\sim 1-5 \times 10^{3}\,{\rm km\,s^{-1}}$, outflow rates up to $\sim100-1000$ times the BH accretion rate, and spatial extents of $\sim0.1-10$\,kpc \citep{tremonti:postsb.outflows,prochaska:qso.outflow,moe:strong.agn.outflow.feedback,wild:postsb.fb.evidence.and.gives.all.ell,fischer:2010.mrk.231,humphrey:2010.type2.qso.feedback,dunn:agn.fb.from.strong.outflows,bautista:2010.strong.agn.fb,feruglio:2010.mrk231.agn.fb,sturm:2011.ulirg.herschel.outflows,rupke:2011.outflow.mrk231,coil:2011.postsb.winds,greene:2011.nlr.outflows,greene:2012.quasar.outflow,faucher-giguere:2012.felobal.model,borguet:2012.bal.outflow.high.energy.kpc.scale,cimatti:2013.agn.outflow.hundreds.kms.common.intermediate.ssfr.highz,cicone:2014.molecular.agn.outflows,harrison:2014.kpc.scale.agn.outflows.common,harrison:2015.agn.outflow.10kpc.driven.by.jets,zakamska:2014.qso.feedback.narrowline.widespread,zakamska:2015.agn.high.vel.narrow.line.outflows}.      In some cases, however, it remains unclear to what extent these outflows are driven by AGN activity vs. star formation. 

The  physics of how AGN-powered outflows interact with the ISM and affect the fueling of the AGN itself -- how inflow and outflow are governed on scales between the small-scale viscous accretion disk ($\ll 0.1\,$pc) and the  galaxy proper ($\gtrsim 0.1$\,kpc) -- remains highly uncertain. There have been many theoretical studies of different ``modes'' of AGN feedback (see references above, as well as \citealt{ostriker:2010.momentum.driving.feedback,ciotti:2010.radiative.mechanical.fb.model,choi:bh.fb.disk.merger.models,choi:agn.fb.massive.elliptical.bal.winds,steinborn:2015.qso.mode.radio.mode.model}, and references therein). However, to date, most of these studies have treated the interstellar medium with relatively simple ``sub-grid'' prescriptions that ignore the additional complications introduced by stellar feedback, interstellar turbulence, and/or the small-scale phase structure of the medium around an AGN. In order to build on these models and model the interaction of AGN outflows and the ISM with greater fidelity, it is critical to include both a realistic description of the physics of the ISM, star formation, and stellar feedback, as well as a plausible description of AGN feedback mechanisms.  

Towards this end,  in this paper we use a suite of numerical simulations to study the interaction of quasar-driven winds and a multi-phase ISM. In a series of papers \citep{hopkins:rad.pressure.sf.fb,hopkins:fb.ism.prop} (hereafter Papers I \& II, respectively), we have developed a new set of numerical methods to explicitly model some of the key processes that shape the multi-phase ISM; the simulations include physically motivated, but still subgrid, treatments of stellar radiation pressure, HII photoionization and photoelectric heating, and the heating, momentum, and mass deposition by supernovae (SNe) and stellar winds. The feedback is tied to the young stars with energetics and time-dependence taken directly from stellar evolution models -- this is particularly important in galactic nuclei, since the dynamical times become shorter than stellar evolution timescales.  In a series of papers \citep{hopkins:2013.fire,muratov:2015.fire.winds,onorbe:2015.fire.cores,van-de-voort:2015.rprocess,ma:2015.fire.escape.fractions,ma:2015.fire.mass.metallicity,faucher-giguere:2014.fire.neutral.hydrogen.absorption}, we showed that, on galactic scales, these models produce a quasi-steady ISM in which molecular clouds form and disperse rapidly, with phase structure, turbulent velocity dispersions, and disk and GMC properties in reasonable agreement with observations. Here, we combine these models with  models for AGN accretion and feedback via both Compton heating and  high-velocity winds from the AGN accretion disk, and examine how various forms of AGN feedback affect black hole accretion, AGN obscuration, and the generation of galaxy-scale outflows. We focus on scales of $\sim0.1-100\,$pc, where the accretion rate onto the black hole is determined, and where AGN-powered winds and Compton heating couple to the ISM. For comparison, the ``radius of influence'' of the BH, inside of which it dominates the potential, is $\sim G\,M_{\rm BH}/\sigma^{3} \sim 5\,$pc in the case we study. This is the first in a new series of papers so we highlight a few of the key results but leave more detailed studies for future work.   

The remainder of this paper is organized as follows.   \S \ref{sec:sims} summarizes our galaxy models and our treatment of radiative cooling, star formation and BH growth, and stellar and AGN feedback.   \S \ref{sec:nofb} summarizes the results of simulations with stellar feedback only, while \S \ref{sec:wfb} compares these results to simulations that include AGN feedback.   \S \ref{sec:discussion} summarizes and discusses our key results.  A series of Appendices contain key technical results.   Appendix A describes our implementation of BH feedback.   Appendix B summarizes the effects of including short timescale variability in the assumed BH accretion rate.  Appendix C describes convergence tests and the effects of using alternate numerical methods.   Appendix D shows that in-shock cooling does not  compromise our results.

\begin{figure*}
    \centering
    \plotside{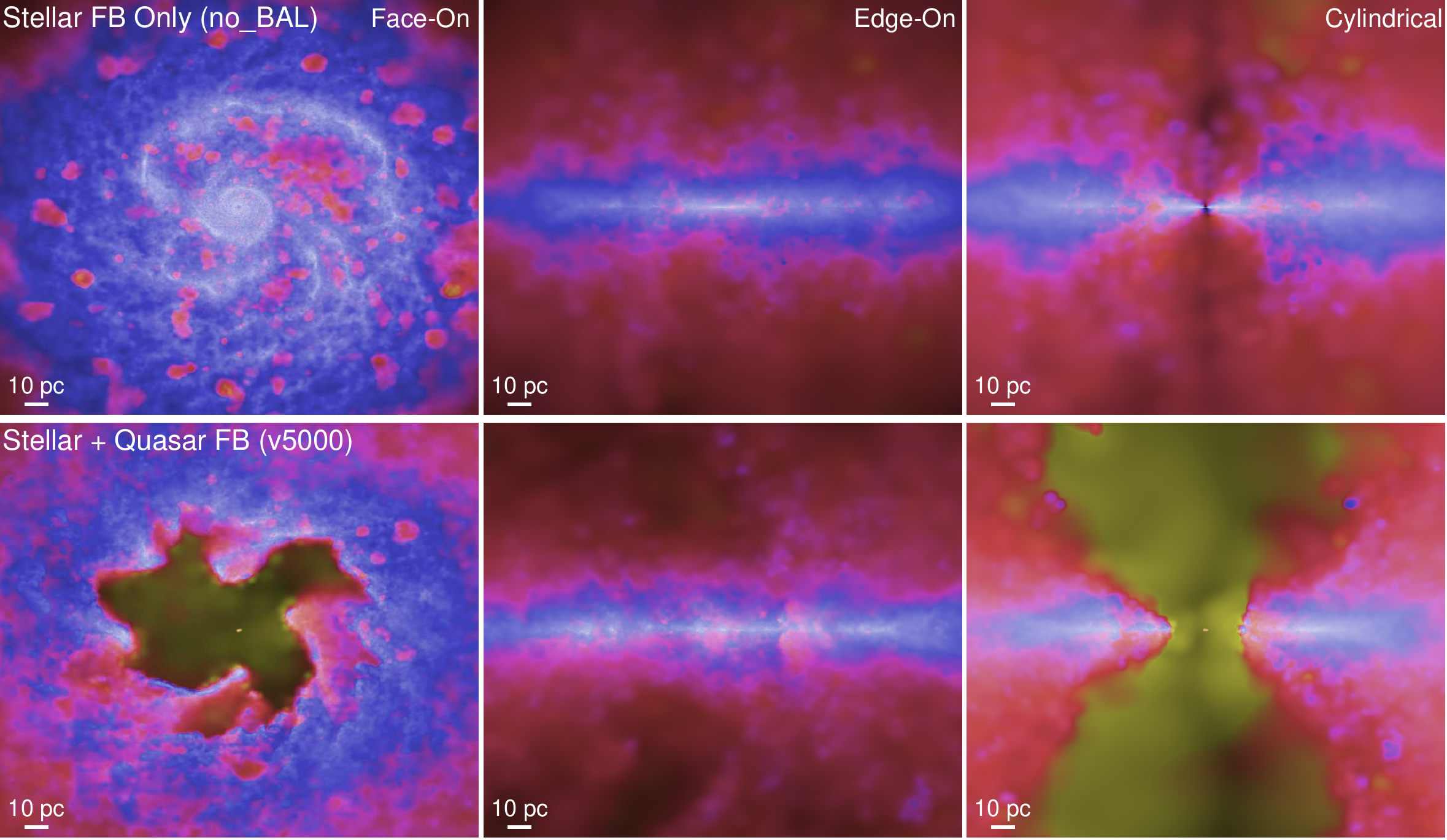}{1}
    \caption{Morphology of the gas in a standard simulation, in face-on ($x,y$; {\em left}), side-on ($x,z$; {\em middle}), and cylindrical ($R,z$; {\em right}) projections.
    The time ($\approx 3\,$Myr since the beginning of the simulation) is $\approx 150$ ($8$) orbital periods at $1\,$pc ($10$\,pc). Brightness encodes projected gas density (increasing with density; logarithmically scaled with a $\approx6\,$dex stretch); color encodes gas temperature with blue material being $T\lesssim1000\,$K molecular gas, pink $\sim10^{4}-10^{5}$\,K warm ionized gas, and yellow $\gtrsim10^{6}\,$K hot gas. 
    {\em Top:} Simulation with stellar, but no AGN feedback ({\bf no\_BAL} in Table~\ref{tbl:sim.param}). A multiphase disk forms; it is mostly molecular inside the central $\sim200\,$pc, with heating by HII regions very localized to small ionized ``bubbles'' and heating by SNe restricted to low-density regions where it can vent vertically. The central $\sim10$\,pc develops a stellar+gas accretion disk dominated by $m=1$ modes. 
    {\em Bottom:} Same, with broad-absorption line winds ({\bf v5000}). The winds blow out a polar cavity and generate an expanding shell in-plane, with occasional dense clumps sinking through to the center. Feedback eventually evacuates the entire nuclear region. 
    \label{fig:morph}}
\end{figure*}

\begin{footnotesize}
\ctable[
  caption={{\normalsize Simulations}\label{tbl:sim.param}},center]{lccccl}{
\tnote[ ]{Parameters describing the simulations in the text: Each employs a gas particle mass of $13.5\,h^{-1}\,\msun$ and minimum SPH smoothing length of $0.0014\,h^{-1}\,$pc. Additional simulations for numerical tests are in Appendix~\ref{sec:appendix:sphmethod}. \\ 
{\bf (1)} Model name \\
{\bf (2)} $\eta_{p}$: Momentum-loading of BAL wind feedback ($\dot{p}=\eta_{p}\,L/c$) \\
{\bf (3)} $\eta_{E}$: Energy-loading of BAL wind feedback ($\dot{E}=\eta_{E}\,L$) \\
{\bf (4)} $\beta$: Mass-loading $\beta\equiv\dot{M}_{\rm BAL}/\dot{M}_{\rm BH}$ (determined by $\eta_{p}$ \&\ $\eta_{E}$) \\ 
{\bf (5)} $v_{\rm BAL}$: AGN wind launching velocity at the simulation resolution (in ${\rm km\,s^{-1}}$; determined by $\eta_{p}$ \&\ $\eta_{E}$)
}
}{
\hline\hline
\multicolumn{1}{c}{Model} &
\multicolumn{1}{c}{$\eta_{p}$} &
\multicolumn{1}{c}{$\eta_{E}$} &
\multicolumn{1}{c}{$\beta$} &
\multicolumn{1}{c}{$v_{\rm BAL}$} & 
\multicolumn{1}{c}{Notes} \\
\hline
{\bf no\_BAL} & 0 & 0 & 0 & 0 & {no AGN FB} \\ 
{\bf v5000} & 1 & 0.008 & 6.0 & 5,000 & {``default''} \\ 
{\bf v5000\_hiP} & 10 & 0.08 & 60 & 5,000 & {high-momentum} \\ 
{\bf v5000\_loP} & 0.1 & 0.0008 & 0.6 & 5,000 & {low-momentum} \\ 
{\bf v30000} & 1 & 0.05 & 1.0 & 30,000 & {high-energy} \\ 
{\bf v500} & 1 & 0.0008 & 60 & 500 & {low-energy} \\
{\bf v5000\_C} & 1 & 0.008 & 6.0 & 5,000 & {+Compton heating} \\ 
{\bf v5000\_iso} & 1 & 0.008 & 6.0 & 5,000 & {isotropic winds} \\ 
\hline\hline\\
}
\end{footnotesize}

\vspace{-0.5cm}
\section{The Simulations}
\label{sec:sims}

The simulations were performed using the {\small GIZMO} code \citep{hopkins:gizmo}. {\small GIZMO} is a multi-method code which can be run with any of several hydro solvers; here we run the code in its smoothed-particle hydrodynamics (SPH) mode, specifically in the ``pressure-entropy'' (``P-SPH'') form which includes several improvements relative to older SPH implementations. Specifically, this is a heavily modified version of the parallel TreeSPH code {\small GADGET-3} \citep{springel:gadget}, in a fully conservative formulation \citep{springel:entropy} which is also density-independent in a manner that allows contact discontinuities and improved fluid mixing \citep[][see Appendix~\ref{sec:appendix:sphmethod}]{hopkins:lagrangian.pressure.sph}. The artificial viscosity, adaptive timestepping, and smoothing kernel are updated following \citet{hopkins:lagrangian.pressure.sph}. The galaxy models and the treatment of star formation and stellar feedback are described in detail in \paperone\ (Sec.~2 \&\ Tables~1-3) and \papertwo\ (Sec.~2). We briefly summarize the salient properties here. 

\begin{figure}
    \centering
    \plotone{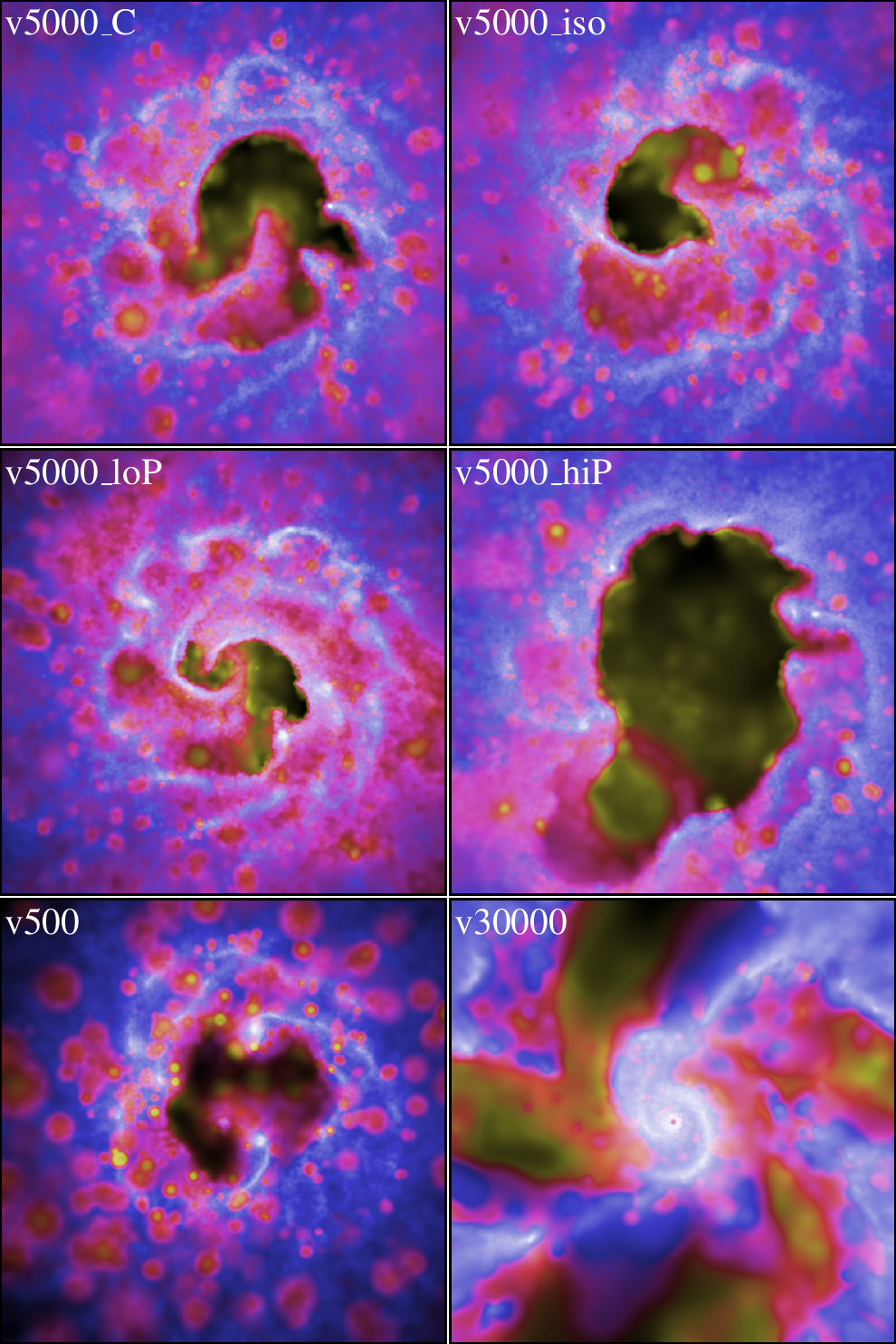}{0.94}
    \caption{Face-on morphology of the gas in the additional simulations from Table~\ref{tbl:sim.param}, as Fig.~\ref{fig:morph}, at the same time (and same scale). {\em Top:} Compton heating ({\bf v5000\_C}) and isotropic vs.\ disk-planar winds ({\bf v5000\_iso}) have no visible effects (the differences are consistent with stochastic variations run-to-run). {\em Middle:} Lowering ({\bf v5000\_loP}) or raising ({\bf v5000\_hiP}) the momentum-loading of the winds leads to smaller/larger bubbles after the initial accretion event, although these in turn alter the subsequent accretion rate (see \S~\ref{sec:wfb}). {\em Bottom:} Lowering the wind velocity (at fixed momentum-loading; {\bf v500}) has no significant effect (although the shocked gas is colder, as expected). Raising the wind velocity to $30,000\,{\rm km\,s^{-1}}$ ({\bf v30000}) creates the most hot gas (as expected); however, this gas appears to mostly vent out from the central regions, so a nuclear disk similar to the no-feedback case re-forms within $\sim 10\,$pc which drives a higher accretion rate at later times.
    \label{fig:morph.all}}
\end{figure}

\vspace{-0.5cm}
\subsection{Initial Conditions}

The initial conditions are a gas-rich nuclear disk in a massive galaxy, drawn from  the large parameter survey of \citet{hopkins:zoom.sims}.
We consider a BH (initial $M_{\rm BH}=3\times10^{7}\,\msun$) in a \citet{hernquist:profile} stellar bulge ($M_{\rm bulge}=10^{10}\,\msun$, isotropic orbits and scale-length $a=1.7\,$kpc) and halo ($M_{\rm halo}=2\times10^{12}\,\msun$, with virial radius, concentration, and velocity appropriate at $z=0$). The BH is surrounded by an exponential nuclear disk of gas and stars (scale-lengths $h_{g}=25\,$pc and $h_{\ast}=10\,$pc, $M_{g}=8\times10^{7}\,\msun$ and $M_{\ast}=2.6\times10^{7}\,\msun$, respectively; stellar disk with vertical sech$^{2}$ profile and dispersions such that $Q=1$, gas disk initially thermally supported with $h/R=0.2$).   The initial surface densities of the gas and stellar disk are thus $\sim 10^5$ $M_{\odot}$ pc$^{-2} \simeq 10$ g cm$^{-2}$.

This is chosen, based on the survey in \citet{hopkins:zoom.sims}, to provide a ``best case'' for extremely rapid BH growth and quasar-level fueling. It is motivated by large-scale simulations of major mergers which produce dense, torus-like structures and high accretion rates \citep{hopkins:m31.disk,hopkins:cusp.slopes,hopkins:inflow.analytics,hopkins:torus,hopkins:agn.alignment,hopkins:sb.agn.delay}, as well as at least some observations indicating the presence of powerful AGN in nuclear starburst ``cusps'' even in galaxies which may not be experiencing extended star formation \citep{diamondstanic:2012.agn.fueling.in.starburst.cusps,mushotzky:2014.agn.present.in.high.sfr.cusps,alatalo:2015.ngc.1266.outflow.suppressing.sf.via.turbulent.injection}. 


The initial gas disk contains $\approx0.6\times10^{7}$ particles; the initial gas particle mass is $\approx20\,\msun$.  We consider a limited resolution comparison in Appendix~\ref{sec:appendix:sphmethod}. The force softening for the BH, gas, and star particles is set to $\epsilon=0.02$\,pc, with minimum SPH smoothing length $=0.1$ times this. We note that all simulations employ the more sophisticated formulation of artificial viscosity described in \citet{morris:1997.sph.viscosity.switch}, which greatly reduces numerical dissipation away from shocks relative to earlier implementations \citep[see e.g.][]{rosswog:2000.sph.switch.viscosity.mod,price:2008.sph.contact.discontinuities}.

\vspace{-0.5cm}
\subsection{Cooling, Star Formation, \&\ Stellar Feedback}
\label{sec:sims:sf}

Gas follows an atomic cooling curve with additional fine-structure cooling to $10$\,K. 
Metal-line cooling is followed species-by-species for 11 tracked species as in \citet{wiersma:2009.coolingtables,wiersma:2009.enrichment}. The enrichment for each species is followed with the time dependent  metal flux directly attached to the mass, momentum and energy flux from stellar winds and SNe Types Ia \&\ II \citep[see][]{hopkins:stellar.fb.winds,hopkins:2013.merger.sb.fb.winds}.

Star formation is allowed only in dense, molecular, self-gravitating regions above $n>10^{4}\,{\rm cm^{-3}}$. We follow \citet{krumholz:2011.molecular.prescription} to calculate the molecular fraction $f_{\rm H_{2}}$ in dense gas as a function of local column density and metallicity, and allow SF only from molecular gas. Following \citet{hopkins:virial.sf}, we also restrict star formation to only gas which is locally self-gravitating, i.e.\ has $\alpha\equiv \delta v^{2}(\delta r)\,\delta r/G\,m_{\rm gas}(<\delta r) \rightarrow (1/4)\,\| \nabla \otimes {\bf v} \|^{2} / (G\,\rho) < 1$  (where the limit is taken as the ``averaging radius'' $\delta r$ vanishes, allowing $\alpha$ to be calculated in a resolution-independent manner only as a function of local properties). Gas which meets all of these criteria forms stars at a rate $\dot{\rho}_{\ast}=\rho_{\rm mol}/t_{\rm ff}$ (i.e.\ $100\%$ efficiency per free-fall time). As shown in \citet{hopkins:virial.sf}, the molecular criterion is not especially important in galaxy centers since most of the dense gas is molecular already, but the self-gravity criterion is important for small scales around black holes, where any simple constant-density threshold for star formation fails to account for the radially-dependent tidal forces. Even in these regions however, the role of these criteria is primarily to determine where stars form; \citet{hopkins:virial.sf} and a number of other studies have shown that the total star formation rate, once fragmentation and stellar feedback are resolved, is set by stellar feedback, and is largely insensitive to details of both cooling and star formation prescriptions \citep[see][]{saitoh:2008.highres.disks.high.sf.thold,hopkins:rad.pressure.sf.fb,hopkins:fb.ism.prop,hopkins:stellar.fb.winds,hopkins:2013.merger.sb.fb.winds,hopkins:stellar.fb.mergers,hopkins:qso.stellar.fb.together,agertz:2013.new.stellar.fb.model}. 

Once stars form, feedback is included in the form of radiation pressure (UV, optical, and IR, allowing for multiple-scattering), stellar winds (fast, young star winds and slow AGB winds), SNe (types Ia and II), photo-ionization and photo-electric heating. Every star particle is treated as a single stellar population with an age based on its formation time and metallicity and mass inherited from its parent gas particle. Feedback includes the relevant mass, metal (with $11$ separately tracked species), momentum, and energy injection to the neighboring gas; all of the relevant quantities (stellar luminosities, spectral shapes, SNe rates, wind mechanical luminosities, yields) for the mechanisms above are tabulated as a function of time directly from the stellar population models in {\small STARBURST99}, assuming a \citet{kroupa:imf} IMF. For every SNe event (or every timestep for winds and single-scattering photon momentum), the relevant energy, momentum, mass, and metals are deposited into the nearest gas particles surrounding each star particle; long-range photo-heating and radiation pressure are treated in a simplified manner assuming spherically symmetric photon propagation from each star particle as an independent source. See \citet{hopkins:rad.pressure.sf.fb,hopkins:fb.ism.prop} for details. The end result of this stellar feedback is a multiphase ISM with a broad range of densities and temperatures.

\vspace{-0.5cm}
\subsection{Black Hole Growth \&\ Feedback}
\label{sec:sims:bhgrowth}

The simulations all include super-massive BHs. The BH is much more massive than the stellar/gas particles, so we do not need to artificially ``force'' the BH particle to stay in the center of the potential, but let it move freely. We cannot, however, directly resolve the viscous accretion disk of the BH on scales $\ll0.1\,$pc. We therefore simply assume that the BH immediately accretes any gas particle gravitationally bound to it (relative velocity less than escape) with apocentric radius (calculated from the particle position and velocity relative to the BH) $<2.8\,\epsilon$ (the minimum Keplerian distance). The rate of particle accretion is capped at the Eddington limit. 

The BH radiates at a luminosity $L=\epsilon_{r}\,\dot{M}_{\rm BH}\,c^{2}$ ($\epsilon_{r}=0.1$ is assumed).\footnote{We also describe in Appendix~\ref{sec:appendix:subgridvar} a model which imposes a spectrum of sub-grid time variability in the accretion rates; however this has no significant effects on the time-averaged results here.} The explicit details of the BH feedback implementation are given in Appendix~\ref{sec:appendix:bhfb}; we briefly summarize them here. Since quasars are believed to have high-velocity, near-planar winds driven off the accretion disk (e.g., \citealt{murray:1995.acc.disk.rad.winds}), we assume that a fraction of the photon momentum drives a wind launched at the resolution scale around the BH from accreted gas. Specifically a fraction of any gas accreted is blown out as a wind with velocity $v_{\rm wind}$, planar with the inflow (by launching particles directly at the accretion radius with this velocity). Two parameters define the wind, the mass-loading and velocity; this is equivalent to specifying the momentum-loading ($\dot{p}_{\rm wind}=\eta_{p}\,L/c$) and energy-loading ($\dot{E}_{\rm wind}=\eta_{E}\,L$) of the wind. Values for the simulation parameters are in Table~\ref{tbl:sim.param}.

We also include Compton heating \&\ cooling from the radiation field. Following \citet{sazonov04:qso.radiative.heating}, this can be approximated with a nearly obscuration-independent Compton temperature of $T_{\rm Compton}\approx2\times10^{7}\,$K. We add the appropriate Compton rates to the standard cooling function (with a limiter following \citealt{cafg:2012.egy.cons.bal.winds} to account for rate-limiting by Coulomb collisions at the high temperatures that can obtain in strong shocks). 

\begin{figure}
    \centering
    \plotone{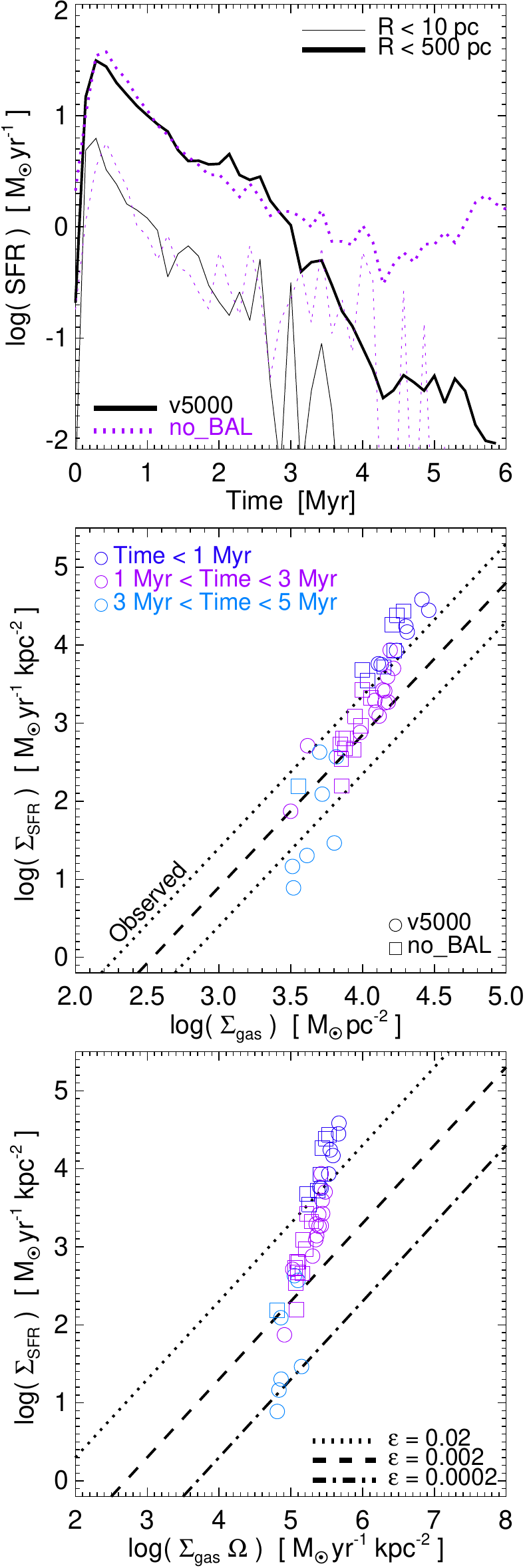}{0.75}
    \caption{{\em Top Panel:}  Star formation rate within 10 and 500 pc regions for simulations with ({\bf v5000}) and without ({\bf no\_BAL}) AGN feedback.  {\em Middle and Bottom Panels:}   Location of the same simulations on the Kennicutt-Schmidt relations at different times.    The star formation rate surface density and gas surface densities are averages within 10 pc and the rotation rate in the bottom panel is also measured at 10 pc.   The observations in the middle panel (dashed line $\pm 0.5$ dex is from \citet{narayanan:2011.xco.model}'s variable $X_{CO}$ model) are based on a range of galaxies, not just galactic nuclei, but nonetheless provide a useful point of comparison.     The star formation efficiency per dynamical time  evolves significantly with time during the simulation, with a relatively high star formation efficiency in the burst of star formation at early times followed by a more prolonged period of lower star formation efficiency.    \label{fig:sfr}}
\end{figure}

\begin{figure}
    \centering
    \plotone{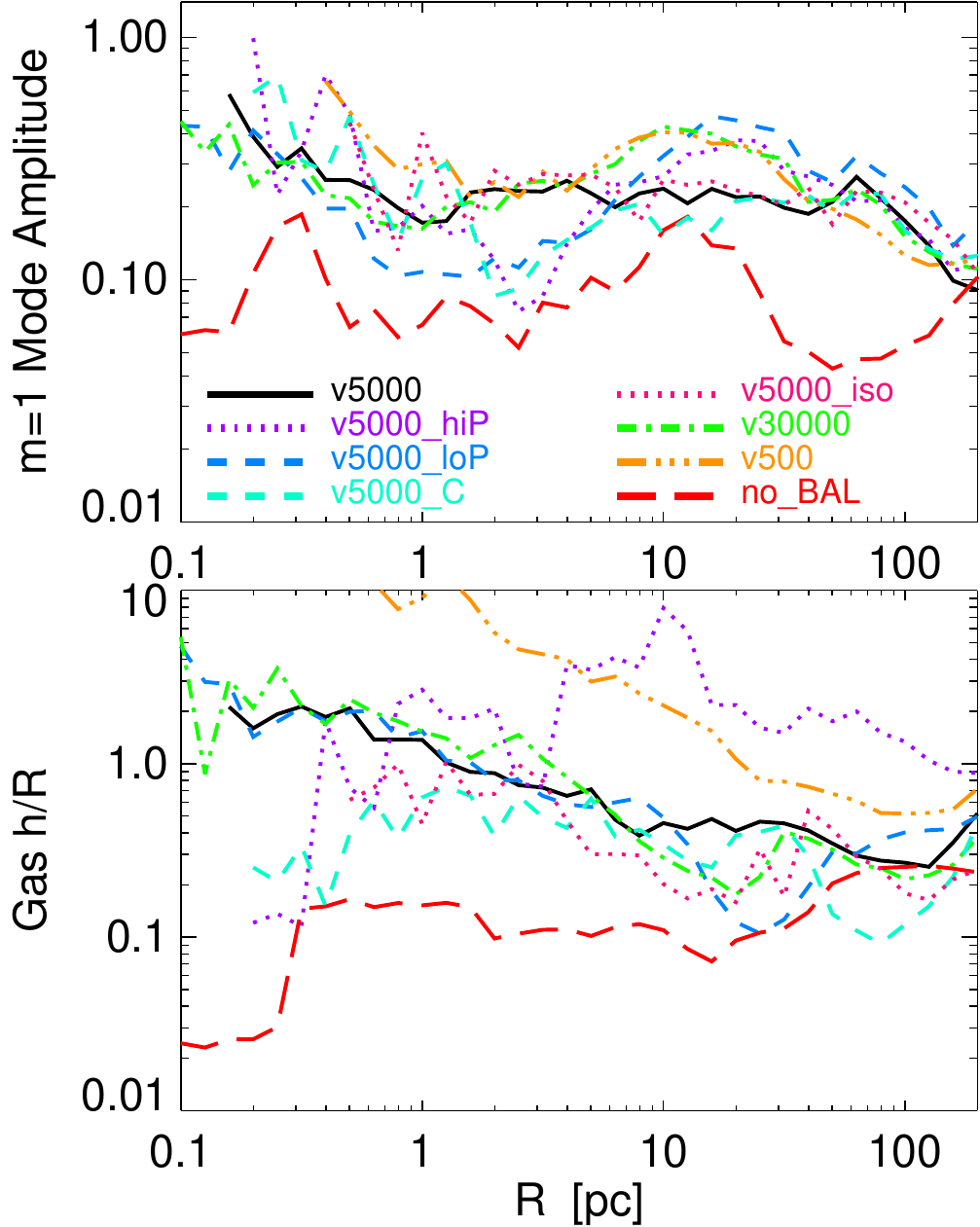}{0.95}
    \caption{Time-averaged structural properties of the simulations. {\em Top:} $m=1$ mode amplitude $|a_{m=1}|$ in the cold molecular gas, as a function of radius. With no BAL feedback the large spiral modes are visible here; with BAL winds the order-unity asymmetries introduced by the AGN wind impacting the ISM dominate. 
    {\em Bottom:} Gaussian disk scale-height ($h/R$) versus radius. With no BAL winds, a modest $h/R\sim0.1-0.2 \sim |a_{m=1}|$ is supported by the combination of stellar feedback and gravitational instabilities. With BAL winds, $h/R$ is greatly enhanced because there is little gas and it is often dominated by escaping/venting polar winds.  Even in the latter case, the scale-height of the cold rotating gas remains modest, similar to that in the non BAL wind simulation (see Fig. \ref{fig:morph}).
    \label{fig:structure}}
\end{figure}

\begin{figure}
    \centering
    \plotone{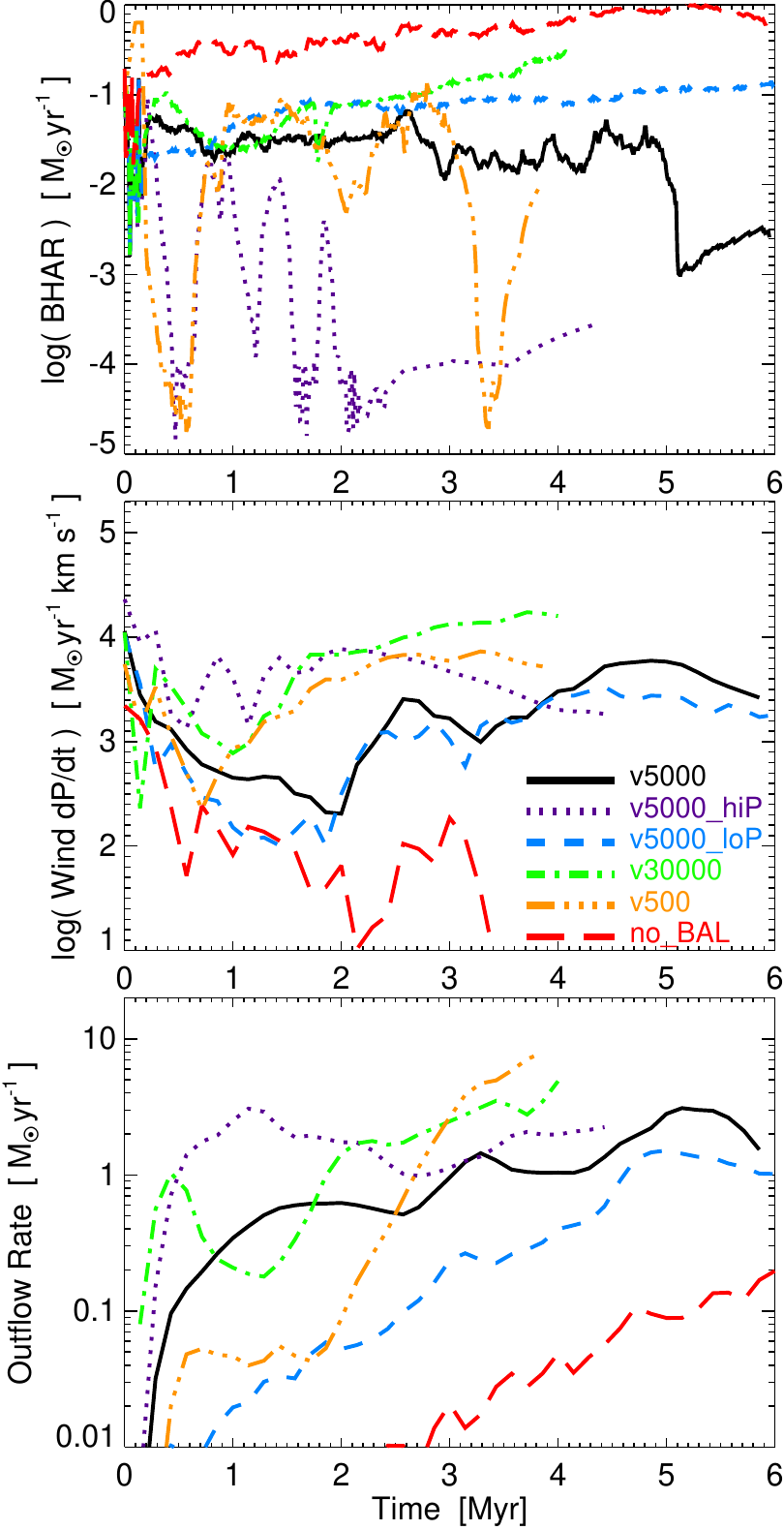}{0.95}
    \caption{{\em Top:} BH accretion rate vs.\ time. To make systematic differences clear we smooth the rates in a $\sim 2\times10^{5}\,$yr window (see Appendix~\ref{sec:appendix:subgridvar} for an un-smoothed case).  AGN feedback suppresses the BH accretion rate relative to simulations without AGN feedback, with higher momentum-loading in the input AGN wind (higher $\eta_{p}$) leading to  lower $\dot{M}_{\rm BH}$. {\em Middle:} Total momentum flux in the galaxy-scale outflow at $>10\,$pc in each model, vs.\ time. The models with BAL winds all equilibrate at broadly similar outflow momentum flux. This explains why higher-$\eta_{p}$ models adjust to have lower $\dot{M}_{\rm BH}$. {\em Bottom:} Corresponding mass-outflow rate of the wind at $>100\,$pc.   Simulations with AGN feedback have dramatically larger outflow rates from the nuclear region relative to the simulation without AGN feedback. 
    \label{fig:inflow}}
\end{figure}

\begin{figure}
    \centering
    \plotone{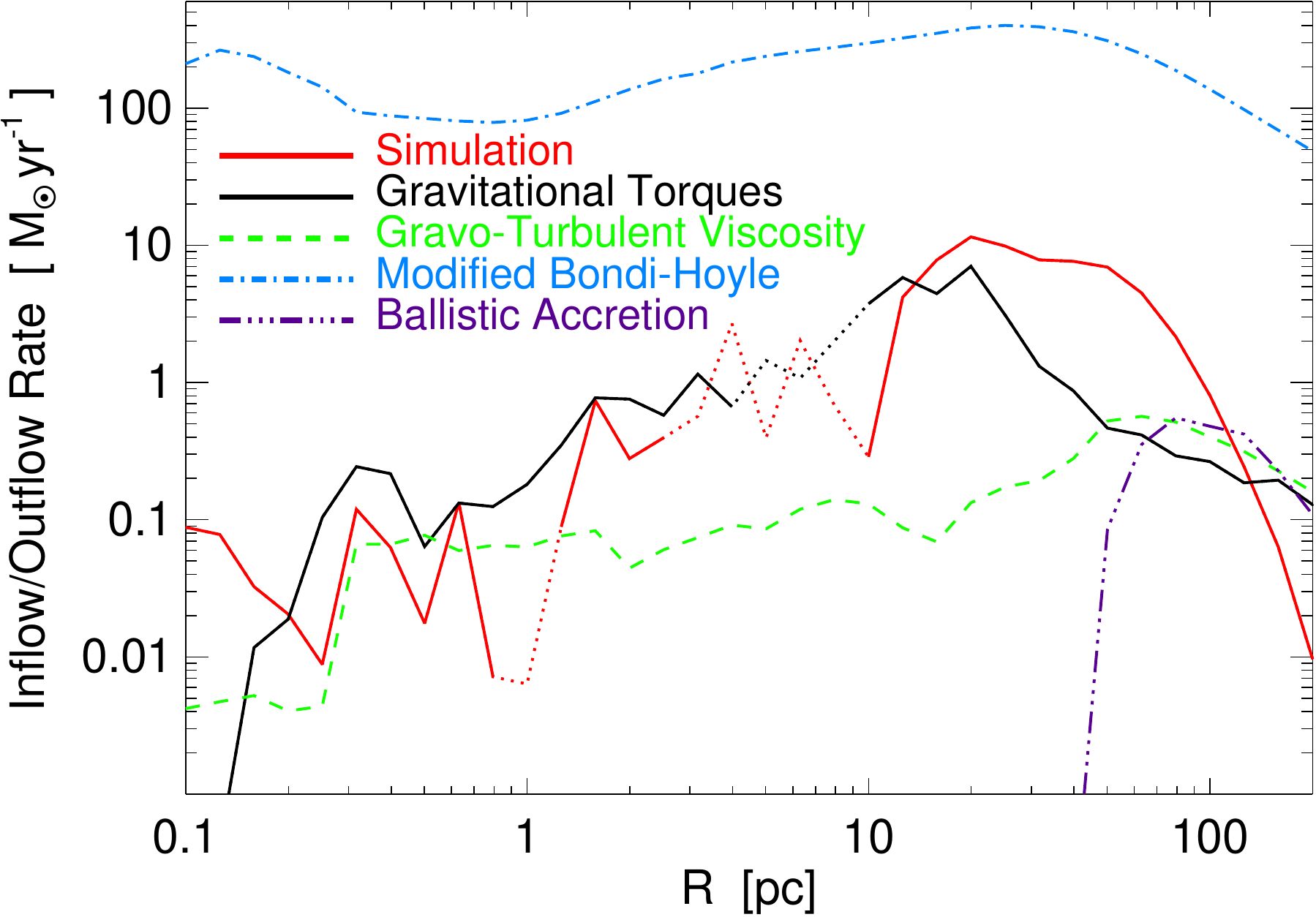}{1.0}
    \caption{Inflow/outflow rate vs.\ radius in the model with stellar feedback alone, averaged over time for a few dynamical times. Negative values (outflow) are dotted, positive values (inflow) are solid (absolute value plotted for the sake of a logarithmic projection).$^{\ref{foot:zerolog}}$ We compare several analytic accretion rate models (\S~\ref{sec:nofb}). The ``gravitational torques'' estimator (resonant exchange between gas+stellar gravitational instabilities) is accurate within a factor $\sim3$ at all radii $<100\,$pc, and correctly predicts the major sign (inflow/outflow) changes. Spherical accretion models fare poorly: the ``pure Bondi'' estimator gives $\sim10^{7}\,\msun\,{\rm yr^{-1}}$, too large to fit on the plot; the ``modified Bondi-Hoyle'' estimator over-predicts by $\sim2-4\,$dex. ``Ballistic accretion'' from turbulence fares poorly in the opposite manner (predicting $\ll 10^{-4}\,\msun\,{\rm yr^{-1}}$ at $R\lesssim 40\,$pc). ``Gravito-turbulent viscosity'' is dimensionally reasonable, but under-estimates $\dot{M}$ by factors of $\sim5-50$ near the BH radius of influence, where gravitational torques are most prominent, and does not capture the sign information. Around $\sim0.1$\,pc, resolution effects from our discrete particle number become important (some predictions drop precisely because the BH is accreting particles).
     \label{fig:inflow.vs.models}}
\end{figure}

\vspace{-0.5cm}
\section{Results with Stellar Feedback, but No Black Hole Feedback}
\label{sec:nofb}


Fig.~\ref{fig:morph} ({\em top row}) shows the morphology of the high-resolution no AGN feedback run at a typical time after a few orbital periods, when the system has reached an approximate statistical steady state (Fig.~\ref{fig:morph.all} compares the additional simulations in Table~\ref{tbl:sim.param}).   Figure \ref{fig:sfr} shows the star formation rate as a function of time for simulations with and without AGN feedback ({\em top panel}) and two versions of the Kennicutt-Schmidt relation describing the star formation law for these nuclear-scale simulations ({\em bottom two panels}).   Note that in the simulation with only stellar feedback, there is an initial burst of star formation but after a few Myr, the star formation rate settles into an approximate steady state at $\dot M_\star \sim 1 \, M_\odot$ yr$^{-1}$ within $\sim 1\,$kpc.   The image in Fig.~\ref{fig:morph} is shown in the latter phase. Within $<10\,$pc, stellar feedback alone does clear most of the gas after a few Myr; this is recycled in a small-scale fountain on a similar timescale. The dynamics of these small-scale burst-quench cycles is explored in more detail in \citet{torrey.2016:fire.galactic.nuclei.star.formation.instability}.  

\subsection{Black Hole Accretion}

Fig.~\ref{fig:morph} shows that the gas disk  exhibits  strong non-linear $m=1$ spiral wave and eccentric/lopsided disk modes, which are visible in spite of the  inhomogeneous structure of the ISM.   Using simulations on similar spatial scales but with a much less realistic model of the ISM, \citet{hopkins:zoom.sims} showed that non-linear $m = 1$ modes generated by stellar-gas interactions dominate the angular momentum transport in galactic nuclei at and inside the BH sphere of influence. However, that study was limited by the assumption that the ISM gas was described by a simple ``effective equation of state'' (i.e.\ no resolved phase structure, winds, or turbulence, simply single-phase gas with a barytropic non-thermal pressure following \citealt{springel:multiphase}). We confirm their result here with a much more realistic ISM model.

Fig.~\ref{fig:structure} ({\em top panel}) plots the $m=1$ mode amplitudes\footnote{Mode amplitudes are measured in the gas surface density as 
\begin{equation}
| a_{m}(R,t) | = \frac{| \int_{0}^{2\pi} \Sigma(R,\,\phi)\,\exp{(i\,m\,\phi)}\,{\rm d}\phi |}
{ \int_{0}^{2\pi} \Sigma(R,\,\phi)\,{\rm d}\phi }\  
\label{eqn:am}
\end{equation}
} versus radius for the simulations with and without AGN feedback.    For the simulation without AGN feedback, the $m = 1$ mode amplitude found here $\sim 0.1$ is similar to that found in \citet{hopkins:zoom.sims}'s simulations with gas fractions $\gtrsim 0.5$.  This suggests that the mode excitation and saturation physics is at least broadly similar in spite of the more dynamic multi-phase ISM present in our  simulations.

Fig.~\ref{fig:inflow} shows the black hole accretion rate,\footnote{To highlight the systematic differences between runs, we boxcar-smooth each curve in a time window of $2\times10^{5}\,$yr. In Appendix~\ref{sec:appendix:subgridvar} we show that there is variability on all resolved timescales in the simulations \citep[as also seen by][]{novak:2011.bh.feedback.cycles,gan:2014.mixed.feedback.models.isolated.elliptical,dubois:2014.bh.pop.evol.cosmo.sims}, and we even consider a model for un-resolved time variability. However, we caution that some of the resolved small-timescale variability is almost certainly artificial here (owing to the assumption that individual gas particles are accreted discretely) -- there are $\sim 10^{4}-10^{5}$ such accretion events per simulation. The BH accretion rate would be likely smoothed out if these were accreted into a viscous disk, which then accretes onto the BH.} the outflow rate from the galactic nucleus, and the total momentum flux in the outflow as a function of time.\footnote{\label{foot:zerolog}The net inflow rate at radius $R$ is given by $\dot{M}=\Delta R^{-1}\,\int {\rm d}M_{\rm gas}\,v_{R}$ in an annulus.   The outflow rate is the same integral, but only over ${\rm d}M_{\rm gas}$ where $v_{R}>0$. The rates are time-averaged in each annulus (which also removes the spurious radial velocity contribution from e.g.\ stationary modes). Because of finite bin-widths the inflow rate can change sign discretely from bin-to-bin.} The simulations clearly find large inflows up to $\sim \msun\,{\rm yr^{-1}}$ to the central $<0.05\,$pc. \citet{hopkins:inflow.analytics} derive an analytic approximation for the inflow rate through each annulus for inflows driven by strong gravitational torques and resonant angular momentum exchange between gas and stars. For modes with complex potential $\Phi_{a}(R)$ and pattern speed $\omega=\Omega_{p}+i\,\gamma$, this is:
\begin{align}
\label{eqn:mdot.full}
\dot{M} & = \Sigma_{\rm gas}\,R^{2}\,\Omega\,{\Bigl |}\frac{\Phi_{a}}{V_{c}^{2}} {\Bigr |}\,
{\Bigl[}\frac{m\,S(\omega,\,\Phi_{a})\,F(\zeta)}{1+{\partial\ln V_{c}}/{\partial \ln R}} {\Bigr]}
\end{align}
with $S(\omega,\,\Phi_{a})$ a phase function and $F$ an order-unity amplitude correction derived in \citet{hopkins:inflow.analytics}, which can  be measured directly in the simulations.
For an $m=1$ mode in a quasi-Keplerian potential, this is approximately $\dot{M}\sim -|a|\,\Sigma_{\rm gas}\,R^{2}\,\Omega$ (with $|a|$ the mode amplitude). 

Fig.~\ref{fig:inflow.vs.models} compares equation~\ref{eqn:mdot.full} to the simulation inflow/outflow rates; the agreement is reasonable, particularly given that the analytic result was derived under the assumption of smooth (non-turbulent) gas flows. Note that divergence between models at $\lesssim 0.1\,$pc is primarily a consequence of our discrete accretion model actually removing particles at this radius, and should be considered with caution.

Fig.~\ref{fig:inflow.vs.models} also compares the inflow rate in our simulations to four alternative proposed accretion rate estimators, none of which does as good a job of reproducing the simulation results. (1) Bondi: $\dot{M}_{\rm Bondi} \approx 4\pi\,G^{2}\,M_{\rm BH}^{2}\,\rho_{\rm gas}\,c_{s}^{-3}$. This over-predicts the accretion rate by an enormous factor $\sim10^{8}$ as most of the gas is cold and molecular, supported not by pressure but by angular momentum. 
(2) Modified Bondi-Hoyle: $\dot{M}_{\rm MBH} \approx 4\pi\,G^{2}\,M_{\rm enc}(<R)^{2}\,\rho_{\rm gas}\,(c_{s}^{2}+\langle V_{\rm gas-bh}^{2} \rangle)^{-3/2}$. This allows for the fact that mass outside the BH itself (e.g. the bulge or nuclear cluster) should, when viewed from gas at large enough distance, act as a point mass in the same way; it also allows for super-sonic relative motion of the gas and BH. This is dimensionally closer to what we measure than [1], but given the low $c_{s}$, it amounts to assuming all gas is in free-fall (neglects angular momentum) -- i.e.\ it is quite similar to simply taking $\dot{M}_{\rm MBH} \sim M_{\rm gas} / t_{\rm free-fall}$ -- and it over-predicts $\dot{M}$ by factors $\sim1000$. Note that replacing $M_{\rm enc}\rightarrow M_{\rm BH}$, as in the standard Bondi-Hoyle formulation \citep[used in][]{springel:models,hopkins:qso.all,hopkins:lifetimes.methods,dimatteo:cosmo.bhs,croft:cosmo.morph.density} only slightly decreases the discrepancy.
(3) Ballistic Accretion: $\dot{M}_{\rm ball} \approx 2\pi\,\Sigma_{\rm gas}\,R^{2}\,\Omega\,(V_{c}/\sigma)\,\exp{(-9\,V_{c}^{2}/16\,\sigma^{2})}$ \citep[this corresponds to accretion of the randomly-populated low-angular momentum ``tail'' of highly turbulent flows from][we generalize their formulae for accretion through each annulus]{hobbs:turbulence.agn.feeding}. This disagrees with the simulations as well; dimensionally it gives $\dot{M}\propto M_{\rm gas}(R)\,\Omega(R)$ but with a ``reduction factor'' $\approx (h/R)^{-1}\,\exp{(-0.56\,[h/R]^{-2})}$, which for $h/R\sim 0.1-0.3$ found here is very small, so that there is very little ballistic accretion.
(4) Gravito-turbulent viscosity: $\dot{M}_{\rm turb} \approx 3\pi\,\alpha\,\sigma_{\rm gas}^{2}\,\Sigma_{\rm gas}\,\Omega^{-1}$ where $\alpha\sim0.005-0.05$ is the (cooling function-dependent) effective turbulent viscosity for a $Q=1$ disk \citep{gammie:2001.cooling.in.keplerian.disks,thompson:rad.pressure,debuhr:momentum.feedback,debuhr:2010.mom.fb.bhgrowth}.\footnote{In Fig.~\ref{fig:inflow.vs.models}, we adopt $\alpha=0.05$, close to the predictions from \citet{gammie:2001.cooling.in.keplerian.disks} given the measured Mach number in the diffuse gas. Changing the value of $\alpha$ will systematically shift the normalization of the predicted curve.} This is dimensionally similar to the gravitational torques scaling but with free-fall slowed by a term $\alpha\,(h/R)^{2}$ instead of $|a|$; over some radii the two are comparable but the former decreases rapidly inside the BH radius of influence (implying accretion would be ``throttled'') while $|a|$ can remain order-unity all the way to the true accretion disk \citep[see][]{tremaine:m31.nuclear.disk.model,bacon:m31.disk,hopkins:inflow.analytics,hopkins:m31.disk,hopkins:slow.modes}. 

The comparisons in this section are based on simulations without AGN feedback.   In the presence of feedback, 
the net accretion rate onto the BH  is determined by a competition between the inflow rate from large scales  set by gravitational torques and the efficiency of AGN feedback at suppressing this inflow in the galactic nucleus.  Our simulations explicitly resolve this competition and produce accretion rates a factor of $\sim 10$ lower than in simulations without AGN feedback (Fig. \ref{fig:inflow}).   For lower resolution galaxy-scale or cosmological simulations it is unclear what the best time averaged accretion rate estimator is to capture this competition between inflow by gravitational torques and AGN feedback; this merits further study in future work.  

\vspace{-0.5cm}
\subsection{Star Formation and Vertical Disk Structure}
\label{sec:vertical}

Fig.~\ref{fig:structure} ({\em top panel}) shows the vertical scale height of the gas disk as a function of radius. The disk is in vertical equilibrium but the dispersions are turbulent (much larger than thermal). As shown in \papertwo\ on larger scales, stars form roughly until feedback can maintain Toomre $Q\approx1$ ($h/R\sim {M_{\rm gas}(<R)}/{M_{\rm enc}(<R)}$) and offset further collapse. At large radii this gives $h/R\sim0.2-0.3$; at $r\sim 3-10\,$pc this is $h/R\sim0.1$ (dispersions $\sim20-70\,{\rm km\,s^{-1}}$).   The cylindrical image in Figure \ref{fig:morph} highlights the modest thickening of the disk at larger radii that is qualitatively analogous to that required in AGN "torus" obscuration models.  As we describe in \S \ref{sec:wfb} this effect is much more dramatic in simulations with AGN feedback because feedback efficiently evacuates the polar region of gas.   

The bottom panels of Figure \ref{fig:sfr} shows our simulations in two common versions of the Kennicutt-Schmidt relation.    The star formation rate surface density and gas surface densities are averages within 10 pc and the rotation rate in the bottom panel is also measured at 10 pc.    The observations in Figure \ref{fig:sfr} are best fits from \citet{narayanan:2011.xco.model} based on a variable $X_{CO}$ factor.   They are shown to provide a point of comparison, but include a range of galaxies, not just galactic nuclei.    The time averaged star formation efficiency in Figure \ref{fig:sfr} is broadly consistent with  observations.   The efficiencies evolve significantly with time, however, with a relatively high star formation efficiency in the burst of star formation at early times followed by a more prolonged period of lower star formation efficiency.   Perhaps most striking is that the star formation efficiency per dynamical time decreases by nearly a factor of $\sim 10$ during the course of the simulations without AGN feedback.   Thus the decline in the star formation rate is not simply due to gas depletion but is also due to the decreasing star formation efficiency.   Note that the duration of the simulation is comparable to the lifetime of massive stars.   Thus the stellar feedback that is effective for most of the duration of the simulation is that due to stellar radiation and stellar winds, since supernovae only start after $\simeq 3$ Myr and have not had significant time to operate.   In addition, because the local dynamical time is short compared to the lifetimes of massive stars, the efficiency of stellar feedback depends primarily on the  surface density of young stars, rather than the star formation rate. We explore the consequences of this for the ``burstiness'' of nuclear star formation and origins of the nuclear-scale Kennicutt-Schmidt relation in a companion paper \citep{torrey.2016:fire.galactic.nuclei.star.formation.instability}.

\vspace{-0.5cm}
\section{Results with Black Hole Feedback}
\label{sec:wfb}

\begin{figure}
    \centering
    \plotone{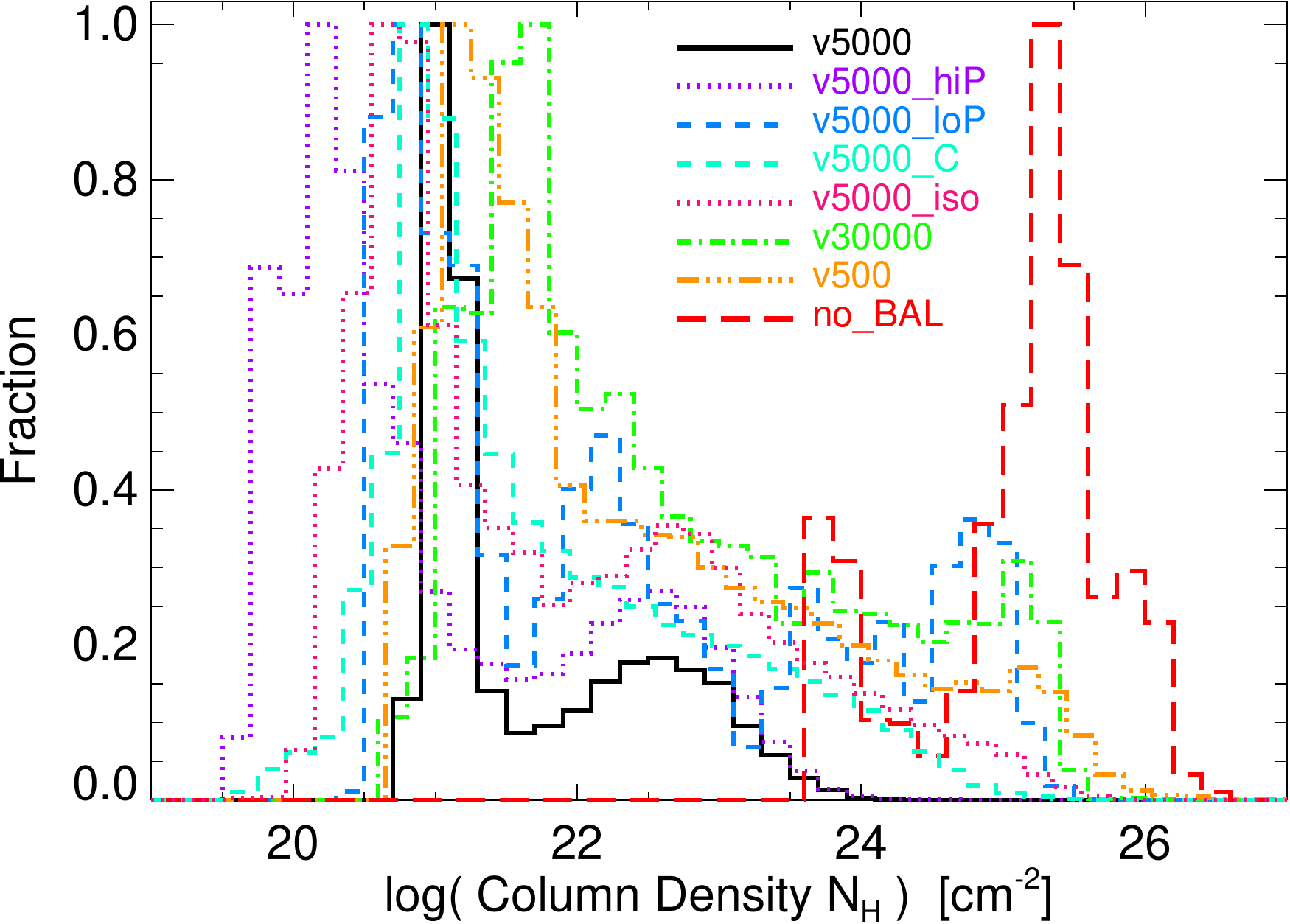}{0.95}
    \caption{Column density distribution on sightlines to the BH in each simulation. We integrate the column along $1000$ sightlines to each BH at each time following \citet{hopkins:lifetimes.letter}, uniformly sampling the sky in solid angle, and show the distribution over all sightlines and times in each simulation.
    Stellar feedback alone  produces a relatively narrow range of very large columns. Simulations with BAL winds have evacuated polar regions with column densities $<10^{22}$ cm$^{-2}$  and overall broader  obscuring column density distributions. The ``clumpy torus'' thus naturally arises from AGN feedback interacting with the large-scale ISM at $\sim10\,$pc.
    \label{fig:nh.dist}}
\end{figure}

\begin{figure}
    \centering
    \plotone{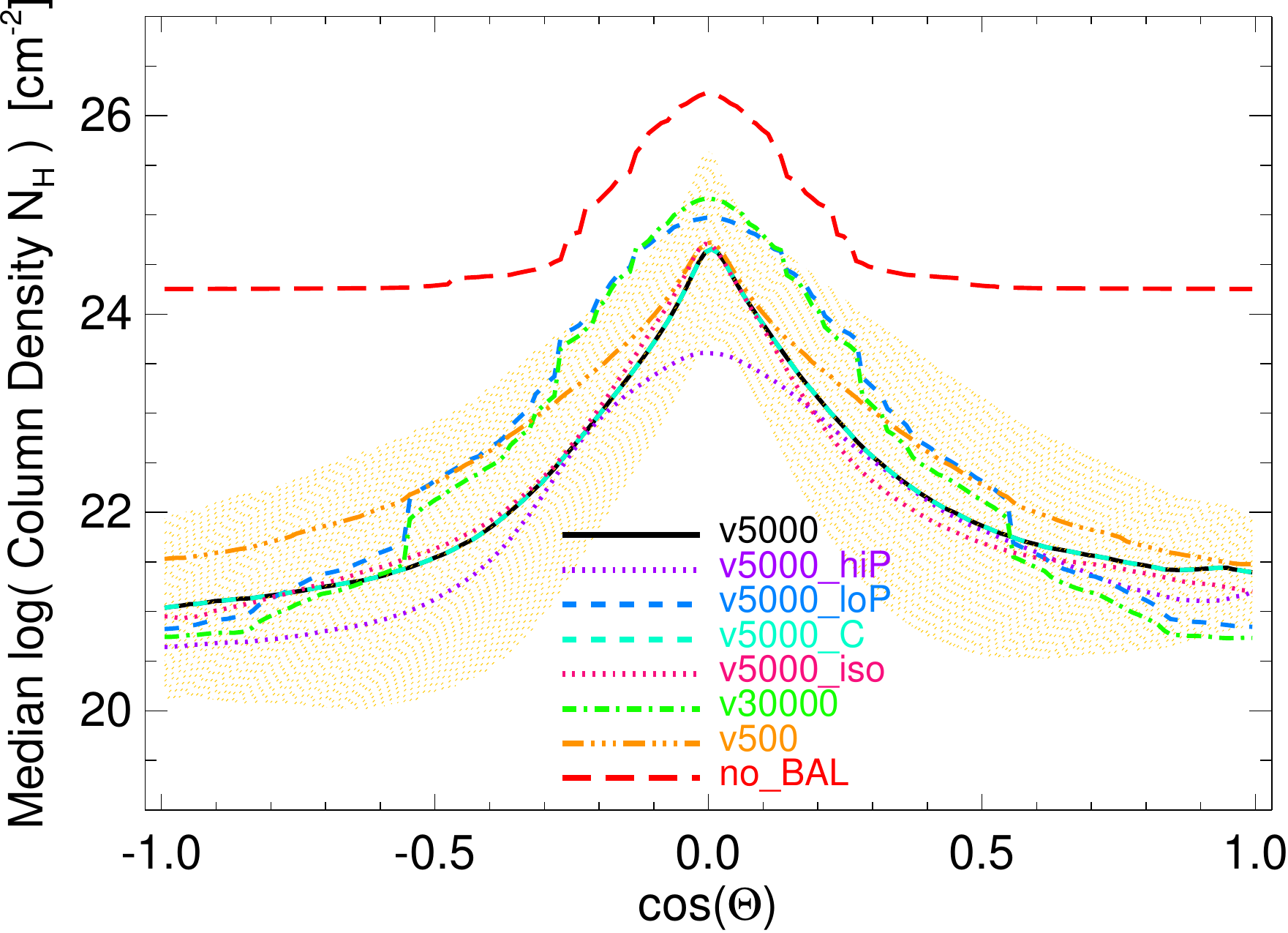}{0.95}
    \caption{Median column density as a function of polar angle $\theta$ (averaged over azimuthal angle $\phi$), calculated over all times as Fig.~\ref{fig:nh.dist}, for the simulations in Table~\ref{tbl:sim.param}. The disk mid-plane ($\cos{\theta}=0$) features Compton-thick columns, as expected, which decline towards the poles ($\cos{\theta}=\pm1$). By clearing the central regions, feedback suppresses the column densities at all $\theta$, but proportionally more towards the poles (where the smaller initial column makes complete evacuation easier) -- this is needed to create optically thin sightlines. In all the simulations, the $1\,\sigma$ scatter in column density at a fixed $\theta$ is $\sim 0.5\,$dex, owing to sub-structure in the gas (Fig.~\ref{fig:morph}). Thus the $\sim95\%$ inclusion range for {\bf v5000} (shaded) is larger than the systematic offset between it and any other simulation with feedback (but not the no-feedback case).
    \label{fig:nh.ang}}
\end{figure}

We now consider the results of simulations with AGN feedback, focusing on the fiducial {\bf v5000} run in which the AGN wind at small radii is injected with $\dot p = L/c$ and $\dot E = 0.008 \, L$. However, we consider variations in both the mass and momentum-loading as well, as outlined in Table~\ref{tbl:sim.param}. 

Fig.~\ref{fig:morph} ({\em bottom}) shows the gas morphology at a few Myr in our {\bf v5000} simulation; there is a clear dramatic impact of feedback on the gas, with the central $\sim$ 30 pc relatively evacuated of gas by the 3 Myr time of these images. Qualitatively, our other simulations with feedback (Fig.~\ref{fig:morph.all}) resemble this case, but with subtle differences discussed below.

The {\bf v5000} outflows  are  launched in the dense disk mid-plane. This drives an expanding shell in the disk plane, with gas  piled up in a narrow ring/shell at the outer (radiative) shock where the winds are encountering the ISM.   This is similar to \citet{cafg:2012.egy.cons.bal.winds}'s models for galaxy-scale winds driven by AGN, though it is not clear if those models quantitatively apply because the hot shocked gas created by the AGN wind is not well-confined -- this may be a limitation of the small scales we are simulating (there is no full galaxy and halo into which the winds can propagate in these simulations).   Indeed, out of the midplane, the entrained mass is modest so outflows coast or are accelerated by hot gas pressure filling the growing central cavity in the disk.

The large impact of the AGN wind on the ambient gas has three closely related effects.   First, it strongly suppresses the star formation in the galactic nucleus, by a factor of $\sim 10-30$ (Fig. \ref{fig:sfr}).   Secondly, it  increases the net outflow rate from the galactic nucleus by a factor of $\sim 10-30$, to $\sim M_\odot$ yr$^{-1}$ (Fig. \ref{fig:inflow}).   Finally,  on longer timescales the BH feedback roughly regulates the BH accretion rate. Specifically, the feedback momentum flux scales as $\dot{p}=\eta_{p}\,L/c$; balancing infall with feedback therefore implies a critical value of $\dot{p}$, so in equilibrium $\langle L \rangle \propto \eta_{p}^{-1}$. This scaling provides a reasonable  approximation over a sufficient time average (Fig. \ref{fig:inflow}), but the evacuation of the central regions clearly leads to very large-amplitude variability on $\sim10^{5-6}$\,yr timescales.

It is useful to directly compare the  momentum flux in the galaxy scale winds (Fig. \ref{fig:inflow}) with those injected at small radii in the AGN wind.   These need not be the same if  AGN feedback produces a bubble of hot gas that does work on the surrounding material, increasing the momentum flux in the wind (e.g., \citealt{cafg:2012.egy.cons.bal.winds}).   For low input wind velocities the momentum fluxes in Fig. \ref{fig:inflow} are comparable to that injected in the AGN wind at small radii, while for higher input wind velocities (in particular, the {\bf v30000} simulation), there is a factor of few boost in the AGN wind momentum flux -- although we inject a momentum flux $\sim L/c$, the outflow momentum fluxes reach $\sim 10\,L/c$, comparable to observed winds on larger-scales which have been decelerated to $\sim 1000\,{\rm km\,s^{-1}}$ \citep{borguet:2012.bal.outflow.high.energy.kpc.scale,cimatti:2013.agn.outflow.hundreds.kms.common.intermediate.ssfr.highz,cicone:2014.molecular.agn.outflows,harrison:2014.kpc.scale.agn.outflows.common,zakamska:2014.qso.feedback.narrowline.widespread,zakamska:2015.agn.high.vel.narrow.line.outflows}. The modest boosts found here are because gas shocked heated by the AGN wind is able to escape relatively easily along the polar direction.  In a more self-consistent calculation, it is possible that the existence of large warps between the disk axis on small scales and that on large scales (e.g., \citealt{hopkins:agn.alignment}) might act to better confine the outflow.  

Note that the outflow rates of $\sim1-10\,M_{\sun}\,{\rm yr^{-1}}$ that we find are relatively modest, compared to many observations at $\gtrsim$\,kpc scales (see references above). This, of course, owes in part to the limited region we are simulating ($\lesssim 100\,$pc) -- there simply is not much material in this volume for the winds to ``sweep up.'' On these scales directly, there are actually relatively few constraints on AGN outflow rates and velocities (most of the constraints above apply either to AGN accretion disk scales, or spatially-resolved outflows on $\sim$kpc scales). However, if we assume the outflow remains momentum-conserving, then for typical galaxy mass profiles we would easily expect it to entrain an order or magnitude or more mass as it propagates from $\lesssim 0.1$\,kpc to a few kpc. 

Figs.~\ref{fig:nh.dist}-\ref{fig:nh.ang} plots the column density distribution and dependence on viewing angle (both averaged over time and in various time intervals) for each of our simulations, both with and without AGN feedback.   The model without AGN feedback predicts virtually no systems with column densities below $10^{24}$ cm$^{-2}$ even along polar sightlines, in stark contrast to observations. Even though the ISM is highly inhomogeneous on larger scales, a small, dense thick-disk or ``halo'' component surrounding the BH in the central $\sim 0.1$\,pc is sufficient to produce these extremely high column densities even in the polar direction. 
The BAL winds have, however, an enormous impact on the column density distribution.  This is not surprising given their impact on the nuclear gas morphology. The polar regions are completely evacuated, giving a large fraction of sightlines that are fully un-obscured. The remaining sightlines follow a broad column density distribution, driven in part by the fragmentation and asymmetries seen in the expanding equatorial shells. 
The evacuation of the central regions out to some radius where $h/R\sim0.1-0.3$ gives a canonical ``torus-like'' global morphology.   This is particularly clear in the cylindrical image shown in the lower right panel of Figure \ref{fig:morph}.

\vspace{-0.5cm}
\subsection{Dependence on the Strength and Form of AGN Feedback}

In Table~\ref{tbl:sim.param}, we outline a series of runs changing the energy and momentum loading of the AGN-driven winds. Here we compare their properties.

First, we add Compton heating/cooling to our ``standard'' case (run {\bf v5000\_C}). Consistent with previous studies \citep{ostriker:2010.momentum.driving.feedback,ciotti:2010.radiative.mechanical.fb.model,choi:2012.bh.fb.idealized,choi:bh.fb.disk.merger.models,choi:agn.fb.massive.elliptical.bal.winds,park:compton.heating.fx.on.bh.growth.important.for.hard.spectra.only} and observational constraints \citep{stern:hot.gas.vs.rad.in.qso.outflows,chatterjee:xray.surface.brightness.constraints.on.agn.feedback}, this has little or no effect on the properties we measure (see Figs.~\ref{fig:morph.all}-\ref{fig:nh.dist}). This should not be surprising: all but the lowest-density gas in the simulations has cooling times much shorter than the Compton heating time. To the extent that Compton heating/cooling is important, it has been speculated that it may be important for cooling in the reverse shock of the BAL winds; however, in the cases we simulate this is not expected to dominate even for a spherical blastwave \citep[see][]{cafg:2012.egy.cons.bal.winds}, and ``venting'' rather than cooling dominates the escape of energy in the hot gas.

We also consider a case ({\bf v5000\_iso}) where the winds from the BH are directed isotropically from the BH, as opposed to in the accretion plane. This is discussed in \S~\ref{sec:appendix:bhfb}; Figs.~\ref{fig:morph.all}-\ref{fig:nh.dist} demonstrate this produces only very small changes (within the range produced by purely stochastic effects) relative to our standard {\bf v5000}. This is also consistent with previous studies \citep{debuhr:2012.bal.plus.radpressure}, and follows simply from the fact that the distribution of accretion directions/angles (the fractional scale-height of accreting gas) is relatively large.

Our {\bf v5000\_hiP} and {\bf v5000\_loP} runs keep the wind velocity fixed but increase/decrease the mass (and momentum) loading by an order of magnitude, respectively, relative to the BH accretion rate. Not surprisingly, a lower (higher) mass-loading produces initially weaker (stronger) outflows: the ``bubble'' in Fig.~\ref{fig:morph.all} is smaller (larger), and this produces slightly higher (lower) column density sightlines in Fig.~\ref{fig:nh.dist}. This has relatively weak effects on the gas properties outside the region being evacuated (Fig.~\ref{fig:structure}). The SFR and Kennicutt-Schmidt relation are essentially indistinguishable from those in Fig.~\ref{fig:sfr} for our {\bf v5000} run, except higher/lower mass-loading translates to earlier/later suppression of the SFR in the immediate vicinity of the BH. The initially stronger outflow (for the same accretion rate) in {\bf v5000\_hiP} suppresses the BH accretion rate substantially in Fig.~\ref{fig:inflow}, while the weaker outflow in {\bf v5000\_loP} allows more rapid growth of the accretion rate compared to {\bf v5000}. Interestingly, this ends up producing a very similar total momentum flux and outflow rate in the winds at $\gtrsim 4\,$Myr. This suggests that indeed we are seeing self-regulation when the BH injects sufficient momentum into the medium to drive outflows that clear its vicinity. This critical asymptotic value is reached eventually in the different runs, and appears to only weakly depend on the feedback parameters. However, how quickly the critical point is reached, and (correspondingly) how much the BH is able to grow, is strongly dependent on the initial feedback mass-loading.

In our {\bf v500} and {\bf v30000} runs, we keep the momentum-loading fixed but vary the initial wind velocity from $500-3\times10^{4}\,{\rm km\,s^{-1}}$. Decreasing the velocity to $500\,{\rm km\,s^{-1}}$ leads to slightly higher average accretion rates in Fig.~\ref{fig:inflow}, but the effect on the circum-nuclear structure (Figs.~\ref{fig:morph.all}-\ref{fig:structure}) and column density distribution (Fig.~\ref{fig:nh.dist}), and outflow rates (Fig.~\ref{fig:inflow}) are weak. This is not surprising since in this limit, the winds are primarily momentum-conserving and so the absolute wind speed (at fixed momentum) does not qualitatively change the dynamics. With $\sim 3\times10^{4}\,{\rm km\,s^{-1}}$ winds, however, we see interesting differences. The faster velocity produces more shock-heated high-temperature gas (unsurprisingly); the lower initial wind mass-loading, however, also appears to lead to more efficient ``venting'' of the wind. This means there is less of a circum-nuclear ``bubble'' carved out in the cold/neutral gas in Fig.~\ref{fig:morph.all}, but rather more pronounced hot gas channels escaping. This in turn allows a thin nuclear disk to re-form (visible morphologically in Fig.~\ref{fig:morph.all}, but also manifest in higher typical column densities in Fig.~\ref{fig:nh.dist}), which then produces higher BH accretion rates seen in Fig.~\ref{fig:inflow}. These higher inflow rates rates lead to a larger net wind momentum injection and outflow rate. Some of these differences are similar to the behavior seen in the ``thermal energy deposition'' models of \citet{choi:bh.fb.disk.merger.models,choi:agn.fb.massive.elliptical.bal.winds}; however, those simulations did not include the physics driving the multiphase structure of the ISM, and so could not capture the full magnitude of ``venting'' effects we see here.

\vspace{-0.5cm}
\section{Discussion}
\label{sec:discussion}

We have used simulations with $< 0.1$ pc resolution to study BH accretion and feedback in gas-rich nuclear disks around massive BHs accreting at quasar-like luminosities.   Our calculations include an explicit treatment of star formation and stellar feedback, which produce a self-consistently inhomogeneous ISM.    We model AGN feedback via Compton heating/cooling and high-speed accretion disk winds injected at small radii.

\vspace{-0.5cm}
\subsection{The Role of Stellar Feedback}

{\em Absent AGN feedback} the properties of the gas disk inside $\sim100\,$pc are as follows. Gas cools efficiently and collapses in a mini-starburst until sufficient young stars are formed to maintain $Q\sim1$ (mostly via radiation pressure-driven turbulence), leading to dispersions $\sim20-100 \ {\rm km\,s^{-1}}$ in a cold nuclear molecular disk. 
As in previous simulations which adopt highly simplified sub-grid models of the ISM \citep{hopkins:zoom.sims}, the disk develops large-amplitude $m=1$ modes in gas and stars, and resonant angular momentum transfer between gas and stellar disks drives rapid inflow of gas, with accretion rates of $\sim 0.1-1 \ M_\odot$ yr$^{-1}$  at $< 0.1$ pc. This agrees well with the analytic \citep{hopkins:inflow.analytics}  predictions for ``gravitational torque''-driven accretion.  In contrast, the Bondi-Hoyle, viscous, or ballistic accretion rate estimators fail to capture the simulation results and are not appropriate for the regimes simulated here, in which much of the gas resides in a rotationally supported (albeit geometrically quite thick) disk (Fig. \ref{fig:inflow.vs.models}).

Stellar feedback does operate somewhat differently in galactic nuclei, as opposed to larger galactic radii, because the local dynamical time $\Omega^{-1}$ is $\lesssim$\,Myr. Young massive stars are sheared into an un-clustered mass distribution (e.g.\ executing hundreds of orbits at $\sim1\,$pc) before they explode.
Rather than local, Jeans-scale clouds evolving independently, we should think of the disk as a coherently evolving, disky ``star cluster'' \citep[see e.g.][]{thompson:rad.pressure}.  On longer timescales, this leads to episodic ``burst-quench'' cycles on small scales, studied in \citet{torrey.2016:fire.galactic.nuclei.star.formation.instability}.

\vspace{-0.5cm}
\subsection{The Role of AGN Feedback}

Our calculations demonstrate that high-velocity winds from the central $\lesssim0.1\,$pc with momentum fluxes $\sim 0.1-1\,L/c$ suggested by observations (e.g., \citealt{cimatti:2013.agn.outflow.hundreds.kms.common.intermediate.ssfr.highz,tombesi:2015.ufo.molecular.outflow.same.galaxy,harrison:2014.kpc.scale.agn.outflows.common,zakamska:2015.agn.high.vel.narrow.line.outflows})  have a dramatic effect on the circum-BH ISM.  In particular, such winds can evacuate gas from the circum-BH disk  (see Figs.~\ref{fig:morph}-\ref{fig:morph.all}).   This suppresses the star formation rate and black hole accretion rate in the galactic nucleus by a factor of $\sim 10$ and enhances the gas outflow rate at $\sim 100$ pc by a comparable factor (Figs. \ref{fig:sfr} \& \ref{fig:inflow}), also similar to observations in at least some systems with powerful on-going outflows \citep{shimizu:2015.agn.specific.sfr.anti.correlated.lowz,guillard:2015.agn.driven.turbulence.as.sf.suppressant,alatalo:2015.ngc.1266.outflow.suppressing.sf.via.turbulent.injection}. As expected, the amount of BH growth required to produce this level of feedback and evacuate gas from the central regions depends inversely on the mass and momentum-loading of the BH accretion-disk winds. The amount of hot gas generated and its ``venting'' depend on the initial wind velocities. Our simulations thus provide support for models in which luminous AGN significantly disrupt the ISM of their host galaxies, at least on scales $\lesssim 100$ pc.   Our simulations also specifically support the hypothesis that luminous AGN may play a key role in driving galaxy-scale outflows from gas-rich galactic nuclei. 

In the plane of the circum-BH disk, the AGN winds decelerate as material is entrained into expanding rings/shells. In the polar direction, however, the galaxy-scale outflows powered by the AGN retain high velocities ($\sim5-30\times10^{3}\,{\rm km\,s^{-1}}$) as they reach kpc scales; although not isotropic, the opening angle for the high-velocity outflow is large ($\gtrsim2/3$ of the sky). A detailed comparison with observations is outside the scope of this work, because we consider only one initial condition, and do not model the large-scale galaxy beyond $\gg 100\,$pc on which many AGN-driven outflows are observed (although a follow-up study designed to compare in detail with these observations is in progress). However, the broad range of velocities present simultaneously within the same system is consistent with outflows observed outside of accretion disk scales in molecular and ionized gas \citep[see e.g.][]{tombesi:2015.ufo.molecular.outflow.same.galaxy,harrison:2014.kpc.scale.agn.outflows.common,zakamska:2014.qso.feedback.narrowline.widespread,alatalo:2015.mrk.231.outflow.variety.of.speeds.some.above.escape}. Similarly, we identify outflowing material across a broad range of temperatures and spatial scales, from molecular to $>10^{6}\,$K gas, and from scales of $\sim0.1-1000\,$pc. Much of the material in the disk plane is accelerated to low velocities and will not escape, but entrains a large mass (comparable to or larger than starburst-driven winds). At least some energy of shocked AGN winds is converted into work, allowing the outflow momentum to reach $\sim 10\,L/c$ in some cases \citep[again, qualitatively consistent with observations; see][]{borguet:2012.bal.outflow.high.energy.kpc.scale,cimatti:2013.agn.outflow.hundreds.kms.common.intermediate.ssfr.highz,cicone:2014.molecular.agn.outflows,harrison:2014.kpc.scale.agn.outflows.common,zakamska:2014.qso.feedback.narrowline.widespread}. This is dictated largely by the  geometry of the surrounding ISM, rather than the AGN wind at small radii.   In particular, simulations with isotropically directed AGN winds on small scales 
give similar results to our default calculations that utilize primarily planar winds (this highlights that once the AGN wind shocks the gas follows the ``path of least resistance'' in the polar direction independent of exactly how the wind is initially directed).   Understanding whether the outflows we find will be confined or halted by the galactic ISM or will continue to escape out of the galaxy will require galaxy-scale simulations.   It is important to stress that the present calculations are not well-suited for addressing this question because our idealized initial conditions do not have, e.g., a gaseous halo or the nuclear warps/mis-alignments seen in both simulations and observations of galactic nuclei.

In our calculations, AGN-driven outflows also have a dramatic impact on obscuration of the AGN itself. AGN winds  evacuate the polar region to allow a fully un-obscured view of the BH.  
AGN winds thus self-consistently produce a torus-like morphology (see, in particular, the lower right panel of Fig. \ref{fig:morph}).   Quantitatively, we find a broad column density distribution from $\sim10^{22}-10^{26}\,{\rm cm^{-2}}$, in reasonable agreement with observations \citep[e.g.][and references therein]{malizia:integral.obscured.agn.column.dist,treister:compton.thick.fractions,risaliti:seyfert.2.nh.distrib,burlon:2011.xr.column.dist.veryhardxr}. The inhomogeneous nature of the ISM also inevitably introduces large ($\sim1\,$dex) variation in obscuring columns even at roughly fixed polar angle -- similar to observational suggestions of ``clumpy'' torii \citep{risaliti:nh.column.variability,mason:ngc1068.torus.obs,
sanchez:circinus.torus.mass,
nenkova:clumpy.torus.model.2,ramosalmeida:pc.scale.torus.emission,
hoenig:clumpy.torus.modeling,deo:2011.z2.clumpy.torii}. 

We find that Compton heating/cooling from the AGN produces weak effects on these scales, consistent with previous studies \citep{choi:2012.bh.fb.idealized,choi:bh.fb.disk.merger.models,choi:agn.fb.massive.elliptical.bal.winds,park:compton.heating.fx.on.bh.growth.important.for.hard.spectra.only}.

\vspace{-0.5cm}
\subsection{Future Work: Other Scales \&\ Forms of Feedback}

These simulations are a first exploration of the  interaction between AGN and stellar feedback on scales between the BH accretion disk and galaxy. We focused on these scales because they are relatively unexplored and yet critical for understanding BH growth and the impact of AGN winds and radiation on the ISM.  It is, however, also clearly important to extend our models to cover a broader range of spatial scales. On smaller scales, understanding the origin of AGN winds and radiation and their ``escape'' from the accretion disk is critical for setting the magnitude and geometry of AGN feedback on pc scales. On galactic scales, we need to understand how the outflows found here interact with the galaxy ISM on long timescales: in particular how this changes galaxy star formation histories and regulates future episodes of BH inflow. This is necessary to determine the effects of feedback on  BH-host galaxy correlations. Also, many observations of galaxy-scale, AGN-driven winds suggest large momentum-loading in the winds, with $\dot{p}$ up to $\sim 10\,L/c$ \citep[see][]{sturm:2011.ulirg.herschel.outflows,cafg:2012.egy.cons.bal.winds,cicone:2014.molecular.agn.outflows}. It will be particularly interesting to see whether these observations are consistent with a model in which this large momentum flux is generated on accretion disk scales (our {\bf v5000\_hiP} model), e.g., by super-Eddington accretion, or whether they require additional large-scale effects such as confinement (and buildup of a pressure-driven bubble) of radiation or hot shocked gas in the galaxies ISM.   This could, e.g., be produced by misalignment between the nuclear scale disk and the galaxy ISM as a whole.

The two AGN feedback mechanisms we have studied here (fast AGN winds and Compton heating), are by no means exhaustive. In future work, we will extend this to include radiation pressure on both narrow lines and dust, photo-heating, and the effects of relativistic jets, all of which can act directly on gas both on the scales we model here but also on much larger scales up to $\gtrsim100\,$kpc. We will also study the effects of different initial conditions (e.g.\ gas fraction and disk-to-BH mass ratio).

\vspace{-0.7cm}
\acknowledgments 
We thank Todd Thompson for helpful discussions, and the anonymous referee for helpful suggestions. Support for PFH was provided by the Gordon and Betty Moore Foundation through Grant \#776 to the Caltech Moore Center for Theoretical Cosmology and Physics, an Alfred P. Sloan Research Fellowship, NASA ATP Grant NNX14AH35G, and NSF Collaborative Research Grant \#1411920. Numerical calculations were run on the Caltech compute cluster ``Zwicky'' (NSF MRI award \#PHY-0960291) and allocation TG-AST130039 granted by the Extreme Science and Engineering Discovery Environment (XSEDE) supported by the NSF. EQ is supported in part by  NASA ATP Grant 12-ATP12-0183, the David and Lucile Packard Foundation, and a Simons Investigator Award from the Simons Foundation.  CAFG was supported by a fellowship from the Miller Institute for Basic Research in Science, by NASA through Einstein Postdoctoral Fellowship Award PF3-140106 and grant 10-ATP10-0187, by NSF through grant AST-1412836, and by Northwestern University funds. 

\vspace{0.1cm}
\bibliography{/Users/phopkins/Dropbox/Public/ms}


\begin{appendix}

\vspace{-0.5cm}
\section{Black Hole Feedback Implementation}
\label{sec:appendix:bhfb}

\subsection{Broad Absorption Line Quasar Winds}

Bright quasars often have BAL winds with velocities of $\sim1000-30000\,{\rm km\,s^{-1}}$. We model these in the most direct manner possible: when a gas particle is accreted, a fraction $f_{\rm acc}$ is actually assumed to accrete on the BH while the remaining $1-f_{\rm acc}$ is blown out as a BAL wind with velocity $v_{\rm BAL}$. There is both observational \citep{schmidt:1999.polarization.bal.qsos.planar.outflows,ogle:1999.bal.qso.polarization.planar.winds} and theoretical \citep{murray:1995.acc.disk.rad.winds} evidence that BAL winds are approximately planar (in or slightly out of the accretion disk plane).  Assuming that the angular momentum vector of the small-scale accretion disk is correlated with that of the sub-pc accreting material, then this corresponds to directing $v_{\rm BAL}$ along the radial vector ${\bf R}\equiv \bf{r}_{i}-\bf{r}_{\rm BH}$ from the BH. On an accretion event we therefore take (for the accreted gas particle) $m_{i}\rightarrow (1-f_{\rm acc})\,m_{i}$, apply the ``kick'' ${\bf v}_{i}\rightarrow {\bf v}_{i} + v_{\rm BAL}\,\hat{R}$, and hold $u_{i}$ (internal energy per unit mass) constant. In model {\bf v5000\_iso} we instead assign the wind direction randomly.   Previous work has shown that randomly directed winds yield results similar to planar winds, but require somewhat larger wind momentum fluxes to achieve the same feedback on the ambient gas \citep{debuhr:2012.bal.plus.radpressure}.

Two parameters must be chosen: the initial outflow mass loading $\beta\equiv\dot{M}_{\rm BAL}/\dot{M}_{\rm BH} = (1-f_{\rm acc})/f_{\rm acc}$ and velocity $v_{\rm BAL}$; observations and theoretical models suggest values of order $\beta\sim1$, $v_{\rm BAL}\sim10^{4}\,{\rm km\,s^{-1}}$ \citep[see e.g.][and references therein]{moe:strong.agn.outflow.feedback,dunn:agn.fb.from.strong.outflows,hamann:2011.bal.nlr.connection,borguet:2012.bal.outflow.high.energy.kpc.scale}

Equivalently, we can translate these parameters into the wind momentum and energy-loading. Since BAL winds are believed to be driven by line radiation pressure in the accretion disk, the available momentum flux is $\dot{p}=L/c$, where $L=\epsilon_{r}\,\dot{M}_{\rm BH}\,c^{2}$ is the luminosity (with $\epsilon_{r}\approx0.1$ the radiative efficiency). The ``initial'' wind momentum and energy are $\dot{M}_{\rm BAL}\,v_{\rm BAL}$ and $0.5\,\dot{M}_{\rm BAL}\,v_{\rm BAL}^{2}$ respectively; thus the energy and momentum-loading are
\begin{align}
\eta_{P} &\equiv \frac{\dot{p}_{\rm BAL}}{L/c} = \beta\,{\Bigl(}\frac{v_{\rm BAL}}{\epsilon_{r}\,c}{\Bigr)}
\approx \beta\,{\Bigl(}\frac{v_{\rm BAL}}{30,000\,{\rm km\,s^{-1}}}{\Bigr)} \\ 
\eta_{E} &\equiv \frac{\dot{E}_{\rm BAL}}{L} = \frac{\epsilon_{r}}{2}\,\frac{\eta_{P}^{2}}{\beta}
\approx0.05\,\beta\,{\Bigl(}\frac{v_{\rm BAL}}{30,000\,{\rm km\,s^{-1}}}{\Bigr)}^{2}
\end{align}
Note for $\eta_{P}=\beta=1$, we recover the canonical $\eta_{E}\approx0.05$ adopted in previous simulations with purely thermal AGN feedback \citep[e.g.][]{dimatteo:msigma,hopkins:lifetimes.letter}.

\vspace{-0.5cm}
\subsection{Compton Heating/Cooling}

The radiation field of the BH will also Compton heat/cool gas in its vicinity. As discussed in \citet{sazonov04:qso.radiative.heating,sazonov:radiative.feedback}, this effect is nearly independent of obscuration: Compton heating is entirely dominated by photons with energies $\gg 10\,$keV (for which we can usually safely ignore obscuration) and Compton cooling by the bolometric luminosity in lower-energy photons (re-distributed, but not, in integral, altered by obscuration). As such even Compton-thick columns result in factor $<2$ changes in the heating/cooling rates. We therefore neglect obscuration and assume the radiation field is isotropic, so that the X-ray/bolometric flux from the AGN on all particles is given by $F_{X}=L_{X}/4\pi\,r^{2}$, with Compton temperature $\approx2\times10^{7}$\,K as calculated in \citet{sazonov04:qso.radiative.heating} for a broad range of observed QSO SED shapes.\footnote{We propagate this flux through the gravity tree, since it follows an inverse-square law when we can neglect obscuration. This makes it trivial to apply the appropriate flux to arbitrary particle numbers, geometries, and numbers of black holes.} In the cooling function, we add the appropriate Compton heating and cooling terms.\footnote{As is standard, cooling is solved implicitly within this function in the regime where the heating/cooling times are short compared to the particle timesteps.} Although Compton cooling depends explicitly on the free electron fraction, for the photon energies dominating heating (much greater than the ionization energy of hydrogen), we can safely approximate Compton heating of bound electrons as identical to free electrons \citep[see e.g.][]{basko:1974.compton.atmospheres,sunyaev:1996.agn.compton}. 

Finally, as shown in \citet{cafg:2012.egy.cons.bal.winds}, some care is needed at the highest temperatures: if the timescale for Coulomb collisions to transfer energy from ions to electrons is longer than the Compton or free-free cooling time of the electrons, this is the rate-limiting process and a two-temperature plasma develops. We therefore do not allow the Compton+free-free cooling rate to exceed the Coulomb energy transfer rate between ions and electrons calculated for an ion temperature $T$ in the limit where the electrons are efficiently cooling $T_{e}\ll T$ \citep[see][]{spitzer:1962.ionized.gases.book,narayan.yi.95:adaf.self.similarity.outflows}. It is important to note that AGN wind-shocked electrons are generally non-relativistic: either immediately post-shock (where most energy is in protons, with electron temperature $T_{e}\sim T_{p}(m_{e}/m_{p})\sim 1.3\times10^{7}\,{\rm K}\,(v_{\rm shock}/30,000\,{\rm km\,s^{-1}})^{2}$), or in later stages when competition between Compton cooling and Coulomb heating regulates the temperature.

\vspace{-0.5cm}
\section{Variability on Un-Resolved Timescales}
\label{sec:appendix:subgridvar}

\begin{figure}
    \centering
    \plotone{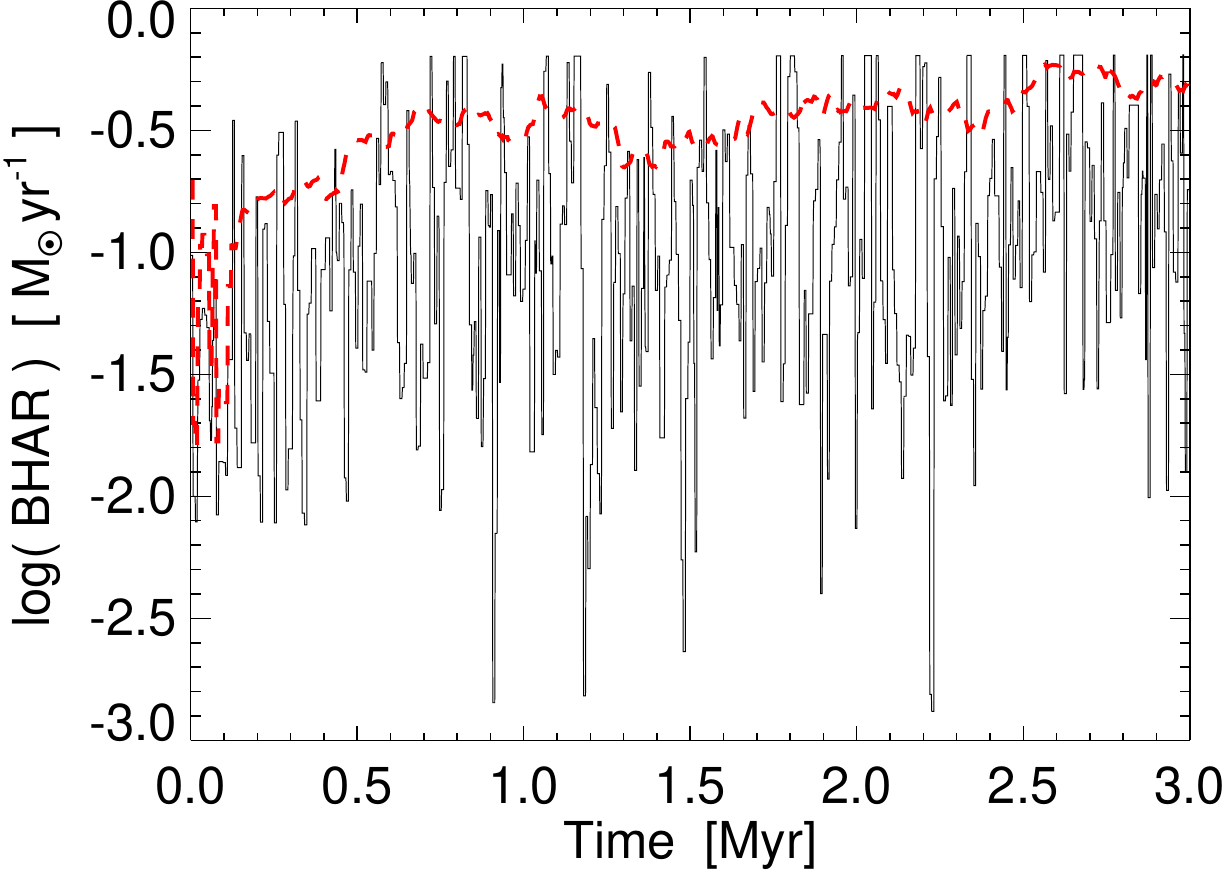}{0.95}
    \caption{BH accretion rate vs.\ time as Fig.~\ref{fig:inflow}, for our ``no BAL'' simulation. We show the rate smoothed in a rolling $2\times10^{5}\,$yr window as in Fig.~\ref{fig:inflow} (thick dashed), and smoothed in a $10^{4}$\,yr window (solid). There is variability on all resolved timescales; however, some of this owes to numerical effects, namely the assumption of instantaneous accretion of discrete gas particles. In Appendix~\ref{sec:appendix:subgridvar}, we describe a model allowing variability even on unresolved (arbitrarily small) timescales.
    \label{fig:tvar}}
\end{figure}

We are able in these simulations to follow inflows to sub-pc scales. This, coupled to the fact that our accretion model is by definition discrete (i.e.\ particles are swallowed individually and {\em instantaneously} accreted onto the BH) leads to large variability on very small timescales, illustrated in Fig.~\ref{fig:tvar}. However, there are still several orders of magnitude between these scales and the BH event horizon, spanned by the Shakura-Sunyaev accretion ($\alpha$) disk. Empirically, AGN exhibit variability on all observed timescales, corresponding to these unresolved spatial scales. Although we cannot resolve these scales, we can make a crude estimate of the effects of this variability by including a sub-grid power-spectrum of luminosity fluctuations and integrating over this to obtain the (modified) momentum flux in every resolved simulation timestep. We quantitatively implement this following the prescription in \citet{hopkins:inflow.analytics}, integrating over a power spectrum with equal logarithmic power per logarithmic time interval, from the minimum resolved timestep down to the orbital time at the innermost stable circular orbit for a non-rotating BH. Performing such an experiment, we find almost no effect on our conclusions. Given the resolved dynamic range in the simulations, this additional variability occurs on extremely small timescales compared to the dynamical times of the outflow -- the timescale over which feedback determines the equilibrium accretion rate. As such, other than adding the chosen random variance to the lightcurve on small timescales, this introduces (relatively) little dynamical effect.

\vspace{-0.5cm}
\section{Numerical Tests}
\label{sec:appendix:sphmethod}

\begin{figure}
    \centering
    \plotone{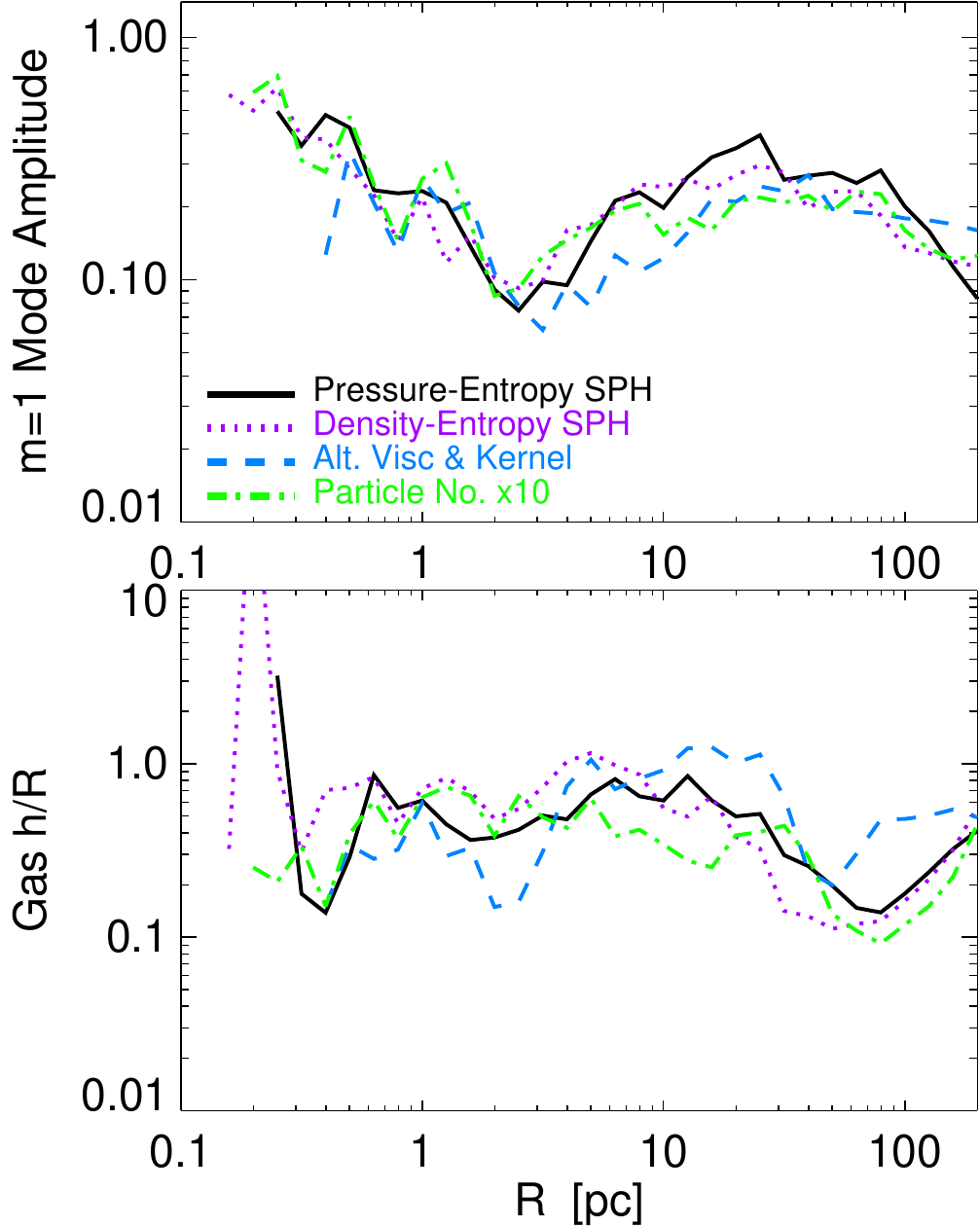}{0.95}
    \caption{Disk structure as Fig.~\ref{fig:structure}, but comparing different numerical methods (see \S~\ref{sec:appendix:sphmethod}: our  ``pressure-entropy'' formulation of SPH developed in \citet{hopkins:lagrangian.pressure.sph}, which performs well in tests of fluid mixing instabilities while maintaining good conservation; a ``density-entropy'' formation (the ``standard'' {\small GADGET} SPH method); a run with a simplified artificial viscosity prescription and SPH smoothing kernel; and a run with the ``pressure-entropy'' formulation but ten times as many particles. The results appear robust to these variations.
    \label{fig:structure.num}}
\end{figure}
\begin{figure}
    \centering
    \plotone{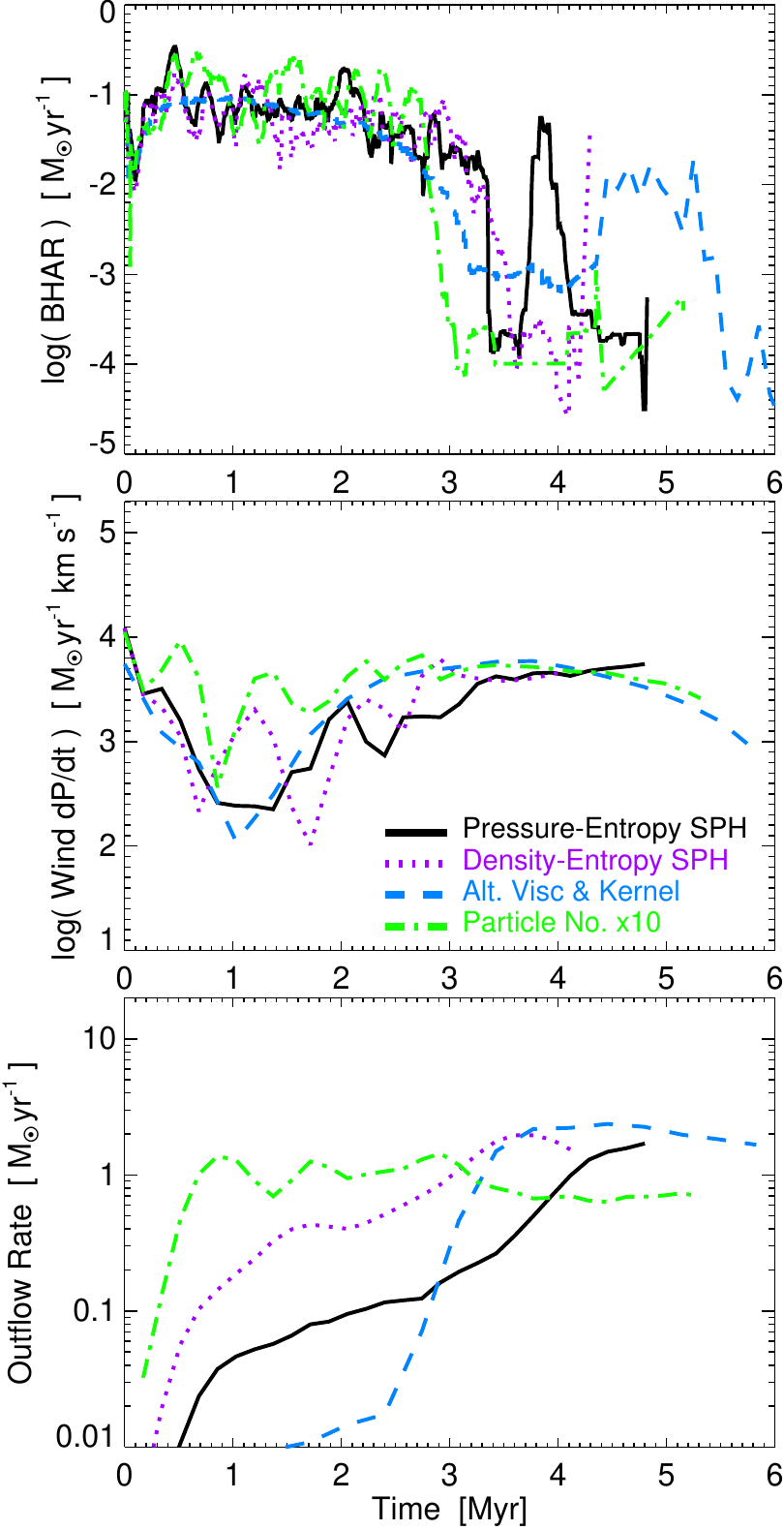}{0.95}
    \caption{Inflow and outflow properties as Fig.~\ref{fig:inflow}, comparing different numerical methods as Fig.~\ref{fig:structure.num}. Again, the results are consistent. Convergence to the asymptotic level of outflow occurs more rapidly at high resolution and more slowly with the modified viscosity and kernel, but this is a result of the non-equilibrium initial conditions. The variations in the late-time accretion rates owe to the chaotic bursts as individual cold gas clumps sink to the center, so we only expect statistical convergence in this stage. 
    \label{fig:inflow.num}}
\end{figure}

We now consider some tests of the robustness of the numerical methods used here. Figs.~\ref{fig:structure.num}-\ref{fig:inflow.num} repeat Figs.~\ref{fig:structure}-\ref{fig:inflow}, but with varied numerical prescriptions. Our default simulations use the ``pressure-entropy'' SPH formulation described in \citet{hopkins:lagrangian.pressure.sph}, which is shown there to give dramatically improved results on in situations with fluid mixing around contact discontinuities (e.g.\ the Kelvin-Helmholtz and Rayleigh-Taylor instabilities) while retaining excellent conservation properties, and includes a number of additional improvements to the treatment of artificial viscosity \citep[see][]{cullen:2010.inviscid.sph}, SPH smoothing kernel accuracy \citep{dehnen.aly:2012.sph.kernels}, and timestep communication relevant for treating extremely high Mach-number shocks \citep{saitoh.makino:2009.timestep.limiter,durier:2012.timestep.limiter}. 

To test whether these subtleties may be strongly influencing our results, we re-run our standard {\bf v5000} simulation instead using a ``density-entropy'' SPH formulation, as in \citet{springel:entropy} (the ``standard'' {\small GADGET} formulation of the SPH equations-of-motion). \footnote{To ensure the simulations are otherwise exactly identical, we have had to re-run the {\bf v5000} simulation in the ``pressure-entropy'' case with a number of small modifications to the algorithm, and on an identical node configuration with pre-set values for certain random number calls. This is done for all tests in this section. For convenience we run the test cases at $1/8$ the particle number in the text.} This produces a ``surface tension'' term at contact discontinuities that suppresses some fluid mixing instabilities, which has been the subject of much discussion in the literature \citep[see][and references therein]{agertz:2007.sph.grid.mixing,read:2012.sph.w.dissipation.switches}. We also re-run the simulations with the pressure-entropy formulation, but adopting the much simpler and more numerically dissipative constant form of artificial viscosity from \citet{gingold.monaghan:1983.artificial.viscosity} (which can significantly alter the behavior in sub-sonic turbulence; see \citealt{price:2011.sph.turb.response}), and a greatly reduced-accuracy SPH smoothing kernel (a 32-neighbor cubic spline, as opposed to our standard 128-neighbor quintic spline). Together these variations produce the range of numerical effects which span the major SPH-grid code differences often discussed in the literature \citep[see references above and][]{price:2010.grid.sph.compare.turbulence,bauer:2011.sph.vs.arepo.shocks,sijacki:2011.gadget.arepo.hydro.tests}. We also run a standard resolution test, increasing the number of particles by a factor of $10$. 

We see very little difference in the results in Figs.~\ref{fig:structure.num}-\ref{fig:inflow.num}. Likewise there is relatively little difference in the column density distributions and gas morphologies. There are some slight differences in the phase diagrams, which correspond to the degree of fluid mixing along phase boundaries (the quantity most affected by these differences), but it is mostly at low temperatures where it does not have significant dynamical effects. This probably relates to the fact that in the numerical comparison studies discussed above, it is generally well-established that different methods agree well in the regimes of super-sonic turbulence and/or systems with dominant external forces. Moreover all of these changes preserve good energy and linear and angular momentum conservation, so to the extent that the outflow is primarily a simple momentum or energy-conserving ``piston,'' and the steady-state stellar feedback is the result of momentum input balancing runaway collapse \citep[see][]{hopkins:rad.pressure.sf.fb,hopkins:fb.ism.prop}, our conclusions should be robust.

In our resolution tests, we see quite good agreement in the BH accretion rate and wind momentum, up to the level of stochastic effects (random differences between simulations). In fact re-running our standard model with different random number seeds for the placement of the initial particles in the disk leads to comparable variations. In the outflow rates, we see the largest differences between runs, at times $<3\,$Myr. These also vary the most dramatically owing to purely random effects; at these times (order one dynamical time at $\sim 100\,$pc), the outflow rate measured at $100\,$pc has just begun to develop, so even small differences in the dynamics can appear significant. The high-resolution case develops a wind more quickly because fragmentation and star formation in the disk at these radii, being better resolved, proceeds faster, making the disk thicker and putting more material into a stellar-feedback driven fountain or wind which the BH wind is then able to entrain. Reassuringly, however, the other simulations soon ``catch up'' as their outer-disk star formation proceeds and more material is entrained, until the long-timescale outflow rates agree well.

\vspace{-0.5cm}
\section{Resolving In-Shock Cooling: Numerical \&\ Resolution Requirements}
\label{sec:inshockcool}

\begin{figure*}
\centerline{\vbox{\hbox{
\includegraphics[width=2.5in]{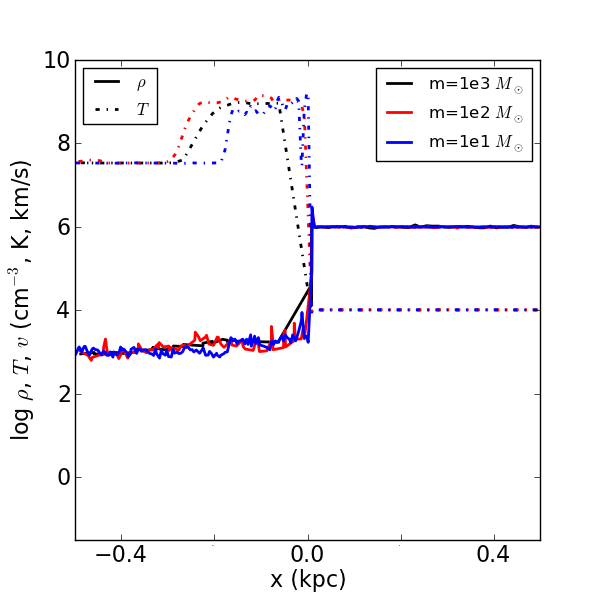}
\includegraphics[width=2.5in]{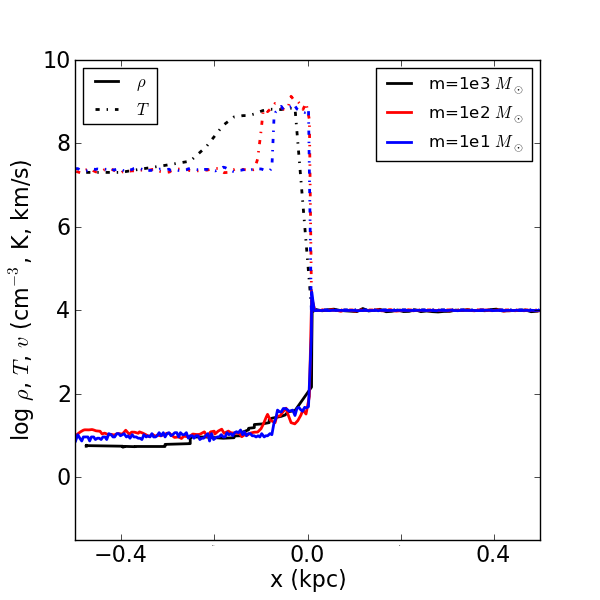}
\includegraphics[width=2.5in]{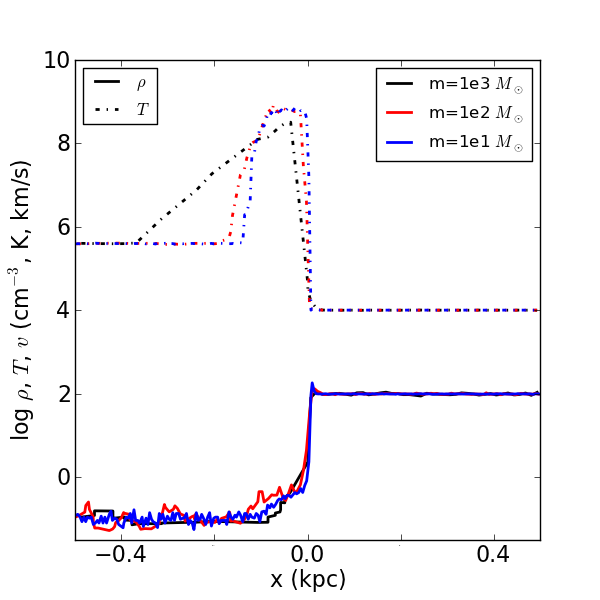}
}}}

\caption{Shock tube tests designed to verify that numerical in-shock cooling does not significantly affect our results at our typical simulation resolution. See Appendix \ref{sec:inshockcool} for details. From left to right, three idealized shock tube problems are shown which represent a BAL wind encountering a cold, dense ISM at radii $R= \{1, 10, 100\} {\rm pc}$, respectively, and evolved to $t=50\,$kyrs. Different colors correspond to the particle masses (as labeled); the highest-resolution case is comparable to our simulations in the main text. At lower resolutions there is noticeable numerical shock-broadening, however the post-shock temperatures are still well-converged (i.e.\ they are not affected by in-shock cooling, which would systematically change the post-shock temperatures at different resolutions).}
\label{fig:ShockProfiles}
\end{figure*}

The SPH hydro solver employed in this work captures the jump conditions associated with shocks over a finite width of order the kernel softening length.   
Gas particles passing through a numerically broadened shock can radiate away energy through traditional cooling channels.
The post-shock gas temperature can therefore be numerically reduced via this ``in-shock cooling" effect when shocks are broadened~\citep[e.g.,][]{hutchings:2000.inshock.cooling,creasey:2011.shock.overcooling}.
In-shock cooling can become significant when the cooling timescale for gas moving through the shock is comparable to the resolution-dependent shock crossing timescale.
For the case of BAL winds with input velocities of $\sim10^4$ km s$^{-1}$, the post-shock gas  is expected to be heated to of order $T\sim 10^9$ K where it will cool inefficiently owning to two-temperature plasma effects \citep{cafg:2012.egy.cons.bal.winds}.
In the absence of efficient cooling channels for the post-shock gas, the outflows will remain energy conserving, efficiently driving outflows via PdV work on the ambient ISM.
As a result, any significant amount of in-shock cooling can impact the post-shock gas temperature, and thus numerically modulate the quasar feedback efficiency studied in this paper.
In this section, we investigate the magnitude of in-shock cooling effects via idealized numerical experiments and find that in-shock cooling should be minimal for the appropriate physical conditions and resolutions used throughout this paper.

In the full feedback simulations presented in this paper, BAL winds are implemented by imparting kicks of 5,000 km s$^{-1}$ to particles near the central black hole based on the accretion rate.
The wind particles shock when they reach the ambient static ISM, thermalizing their kinetic energy, and giving rise to a physical situation similar to that shown in Figure 1 of \citet{cafg:2012.egy.cons.bal.winds}.  
To properly capture the full impact of BAL wind injection on quasar outflows, it is important that the thermalization of the BAL wind kinetic energy at the reverse shock does not suffer substantial from in-shock cooling.
Gas densities, temperatures, and velocities for the reverse shock are set by the pre-shock BAL wind material, which is assumed to be free streaming.
The non-homogenous density structure of the ISM and variability of the AGN radiation field make identifying the impact of in-shock cooling difficult in the full simulations directly.
We instead construct idealized shock tube tests to recreate these conditions in a setting where resolution-dependent in-shock cooling can be directly identified.

We use a three-dimensional shock tube to explore the reverse shock density and temperature profile as a function of physical conditions (i.e. pre-shock density) and numerical resolution.
The idealized initial conditions for the reverse shock include a fast moving medium (imitating the BAL wind material) moving into a static medium (imitating the ambient ISM).
The BAL wind material is given an initial temperature of $T_{{\rm BAL}} = 10^4$ K, however this can change rapidly at the onset of the simulation due to Compton heating off of the AGN radiation if the gas density is sufficiently low.
The BAL material is given a velocity of $v_{{\rm BAL}} = 5,000$ km s$^{-1}$. 
The ambient ISM material is given an initial temperature of $T_{ {\rm ISM }} = 10^2$ K and is initially static.
The initial density for the BAL wind, initial density for the ambient ISM, and incident flux of AGN radiation are dependent on the location of the shock.
We approximate the density of the pre-shock free-streaming BAL wind material to be given by
\[ \rho_{{\rm BAL}} \approx 10^3  {\rm cm }^{-3} \left( \frac{\dot M}{1 \; M_\odot \; {\rm yr}^{-1} } \right) \left( \frac{R}{1 {\rm pc}}  \right)^{-2} \left( \frac{v}{5000 \; {\rm km} \; {\rm sec}^{-1} } \right)    \] 
the density of the ambient ISM as 
\[ \rho \approx 10^6 {\rm cm }^{-3}  \left( \frac{R}{1 {\rm pc}}  \right)^{-2}  \]
and the incident AGN radiation flux as
\[ F_{{\rm AGN}} \approx 10^8 \frac{ {\rm  erg }}{{\rm s \; cm^2 }} \left( \frac{L}{10^{46} {\rm \; erg \; s^{-1}} }\right) \left( \frac{R}{1 \; {\rm pc }} \right)^{-2}. \]
We assume fiducial values of $\dot M = 1M_\odot /{\rm yr}$, $v =5000 \; {\rm km} / {\rm sec}$, and $L =10^{46} {\rm \; erg / sec}$ and run tests for $R= \{1, 10, 100 \}$ pc.
The shock tube uses periodic boundary conditions in a rectangular prism of dimension $1 \times R/(800 \; {\rm pc}) \times R/(800 \;{\rm pc})$ kpc.

Figure~\ref{fig:ShockProfiles} shows the gas density and temperature profiles across the idealized shock at $t=50$ kyrs.  
The three panels show different values for the ambient gas density and AGN radiation flux (corresponding to $R=\{1, 10, 100\}$ pc, as described in the previous paragraph) tests with the legend indicating the gas particle mass resolution in each test.
We find that the lowest resolution test (black line) blurs the location of the reverse shock substantially in the $R=10$ and $R=100$ pc tests.
The two higher resolution tests present with less blurring the of the reverse shock, however there is still an offset present in the location of the reverse shock owing to the low particle number in the pre shock low density BAL wind material.
In terms of in-shock cooling, we find that the post shock gas forms a stable and nearly flat temperature profile which shows little variation as we change the mass/particle resolution for our highest two resolution tests.  
Although some shock broadening is present, the post-shock gas temperatures are not strongly (if at all) impacted by the increasing resolution.  
If in-shock cooling were present,  we would instead expect the post-shock gas temperature to decrease with lower mass resolution.
Since the high resolution tests explored here ($m = 10 M_\odot$) have resolution comparable to that used in the simulations presented in this paper and show little indication in-shock cooling, we conclude that in-shock cooling should not significantly impact the post-shock gas temperatures, and therefor should not have a significant impact on the results presented in this paper.


\end{appendix}

\end{document}